\documentclass[aps,prb,twocolumn,showpacs,preprintnumbers,superscriptaddress,amsmath,amssymb,longbibliography]{revtex4-2}
\usepackage{graphicx}% Include figure files
\usepackage{dcolumn}% Align table columns on decimal point
\usepackage{bm}% bold math
\usepackage{amsfonts}
\usepackage{amsmath}
\usepackage{amssymb}
\usepackage{xcolor}
\usepackage{enumerate}
\usepackage{hyperref}
\usepackage{siunitx}
\DeclareSIUnit{\fu}{{f.u.}}
\usepackage{color}
\usepackage{subcaption}
\usepackage[section]{placeins}
\usepackage{filecontents}

\begin{document}

\title{Octahedral distortions in SrNbO$_3$: Unraveling the structure-property relation}

\author{V. Rosendal}
\affiliation{Department of Energy Conversion and Storage, Technical University of Denmark, 2800 Kgs. Lyngby, Denmark}
\author{W. H. Brito}
\affiliation{Departamento de  F\'{\i}sica, Universidade  Federal de Minas Gerais, C. P. 702, 30123-970, Belo Horizonte, MG, Brazil}
\author{M. Radovic}
\affiliation{Swiss Light Source, Paul Scherrer Institut, CH-5232 Villigen, Switzerland}
\author{A. Chikina}
\affiliation{Swiss Light Source, Paul Scherrer Institut, CH-5232 Villigen, Switzerland}
\author{M. Brandbyge}
\affiliation{Department of Physics, Technical University of Denmark, 2800 Kgs. Lyngby, Denmark}
\author{N. Pryds}
\affiliation{Department of Energy Conversion and Storage, Technical University of Denmark, 2800 Kgs. Lyngby, Denmark}
\author{D. H. Petersen}
\affiliation{Department of Energy Conversion and Storage, Technical University of Denmark, 2800 Kgs. Lyngby, Denmark}

\date{\today}

\begin{abstract}
    Strontium niobate has triggered a lot of interest as a transparent conductor and as a possible realization of a correlated Dirac semi-metal. 
    Using the lattice parameters as a tunable knob, the energy landscape of octahedral tilting was mapped using density functional theory calculations. We find that biaxial compressive strain induces tilting around the out-of-plane axis, while tensile strain induces tilting around the two in-plane axes. The two competing distorted structures for compressive strain show semi-Dirac dispersions above the Fermi level in their electronic structure. 
    Our density functional theory calculations combined with dynamical mean field theory (DFT+DMFT) reveals that dynamical correlations downshift these semi-Dirac like cones towards the Fermi energy.
    More generally, our study reveals that the competition between the \textit{in-phase} and \textit{out-of-phase} tilting in SrNbO$_3$ provides a new degree of freedom which allows for tuning the thermoelectric and optical properties. We show how the tilt angle and mode is reflected in the behavior of the Seebeck coefficient and the plasma frequency, due to changes in the band structure.
\end{abstract}
   
\maketitle

\section{Introduction}
The perovskite (oxide) structure, ABO$_3$, is a versatile structure relevant in many existing and emerging applications~\cite{Bhalla2000} ranging from piezoelectricity~\cite{Panda2015}, thermoelectricity~\cite{Yin2017}, oxygen separation, and solid oxide fuel cells~\cite{Sunarso2017}. It is also a platform  where the coupling of charge, spin, and orbital degrees of freedom takes place, giving rise to numerous materials with interesting electronic and magnetic properties such as superconductivity~\cite{Schooley1964}, colossal magnetoresistance~\cite{Baldini_CMR}, metal-insulator transitions~\cite{Wong2010, Gu2013}, and more recently the realization of correlated Dirac semimetallic states~\cite{Ok2021}.

\begin{figure*}
    \centering
    \includegraphics[width=0.65\textwidth]{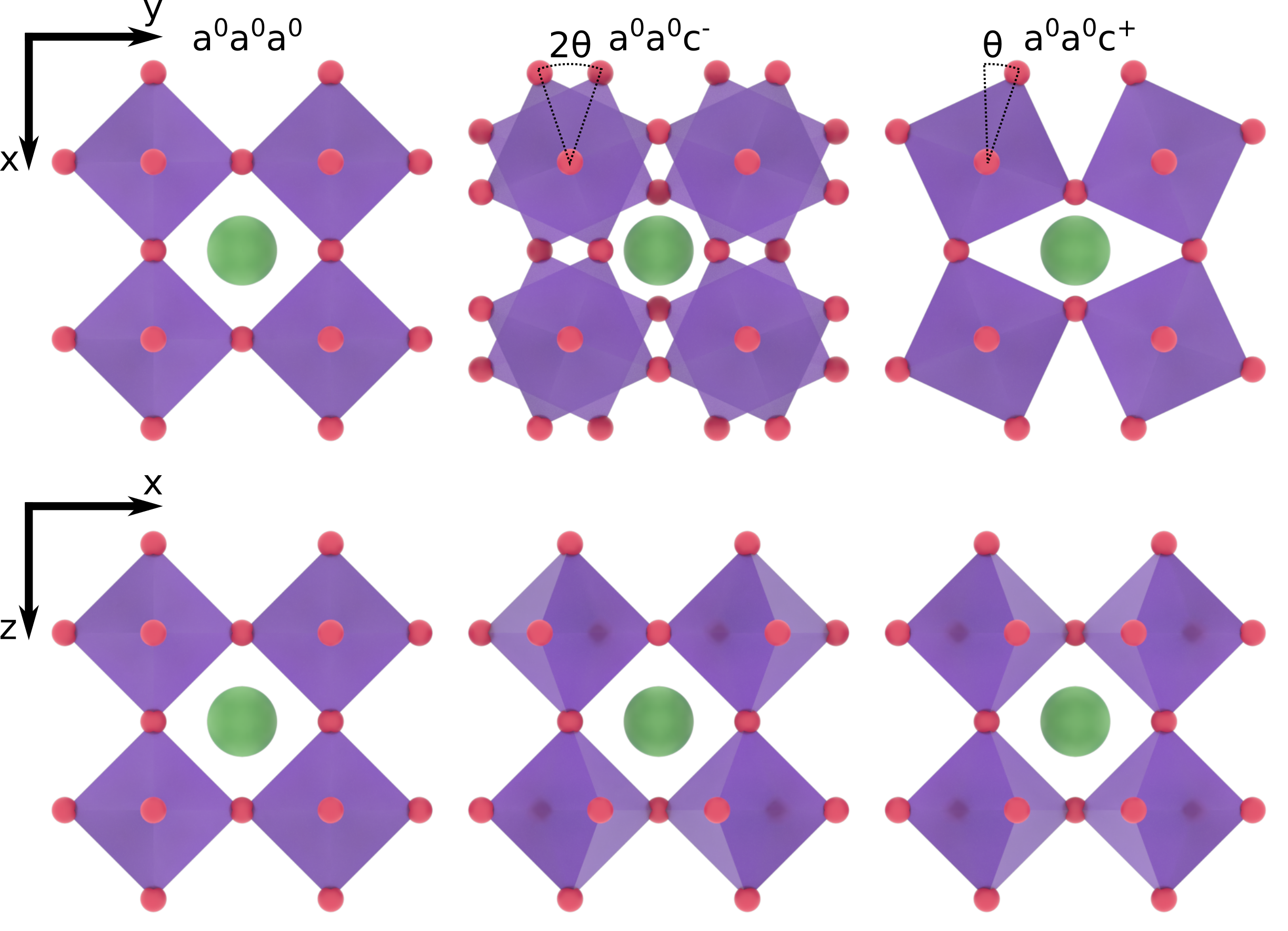}
    \caption{Illustration of different octahedral tilting modes. The left column shows $a^0a^0a^0$ (no tilt), the middle column shows $a^0a^0c^-$ (\textit{out-of-phase} tilting) and the right column shows $a^0a^0c^+$ (\textit{in-phase} tilting). The upper (lower) row shows the structures in the $xy$-plane ($zx$-plane). In SrNbO$_3$ structure,  the octahedra are formed by oxygen (red) centered around niobium atoms. The green spheres correspond to strontium atoms.}
    \label{fig:octahedron}
\end{figure*}

One possible way of changing the properties of perovskite oxides is to apply strain to the crystal.~\cite{Rondinelli2011, Rondinelli2012} This can be achieved by epitaxial growth of films on substrates with different lattice parameters. The lattice mismatch between the substrate and the film will induce strain in the system. The compression (tension) induced by the substrate will shrink (expand) the in-plane B-O bond length from its equilibrium value. As a response, the oxygen ions can displace and alter the B-O-B bond angle and bond lengths, and in this way relieve stress. This displacement will be referred to as \textit{octahedral tilt} or \textit{octahedral rotation}, and it is visualized in Figure \ref{fig:octahedron}.

Strontium niobate, SrNbO$_3$, is a conducting perovskite oxide that has gained interest in recent years.~\cite{Oka2015, Park2020, Ok2021} SrNbO$_3$ has a larger lattice parameter in comparison to the prototypical SrTiO$_3$, i.e., \SI{4.023}{\angstrom} and \SI{3.905}{\angstrom}, respectively.\cite{Oka2015} Park \textit{et al.} investigated SrNbO$_3$ as a potential transparent conductor due to its large gap between the (filled) conduction and valence states.~\cite{Park2020} The energy gap between the valence band maximum and conduction band minimum is ca. \SI{2.3}{\electronvolt}, predicted with density functional theory in Reference \cite{Park2020}. Further interest has been related to the emerging Dirac states in heavily strained SrNbO$_3$ thin films. Both theoretical and experimental work suggests that light electrons emerges in strained SrNbO$_3$ films due to the induced octahedral tilting.~\cite{Ok2021}

Yet, there has been no systematic investigation of the atomic and electronic structure of SrNbO$_3$ under biaxial stress, e.g., found in epitaxial thin films, especially including octahedral tilting. Theoretical calculations provide a fast and unique way to investigate the local distortions, that are difficult to access using experimental techniques.~\cite{Zhao2021} This has been shown to be important for explaining metal-insulator behaviour in $3d$ perovskite oxides using standard density functional theory (DFT) methods.~\cite{Varignon2019}
Distortions such as octahedral tilting alters the electronic configuration, specifically the orbital overlaps and band widths, $W$. The ratio between the interelectronic Coulomb interaction and the band width, $U/W$, describes the strength of the electron correlation. Furthermore, a large ratio can result in a localization of the electrons, i.e. $U/W$ governs the Mott metal-insulator transition.~\cite{Mott2004} Hence, the electrical properties are dependent on the band width, which is connected to distortions such as octahedral tilting.
Similarly, there is a connection between the band width and the optical response of a material. The plasma frequency that governs the optical response can be written as $\omega_p = e\sqrt{n/(m^* \varepsilon)}$, with $e$ being the elemental charge, $n$ is the carrier concentration, $m^*$ is the effective mass and $\varepsilon$ is the permittivity.~\cite{Mirjolet2021} The plasma frequency is then related to the band width through the effective mass $m^* \propto 1/W$. Therefore, a reduction of the band width (or increase in effective mass) results in a redshift of the plasma frequency. Due to the connection between atomic structure and electrical and optical properties discussed here, it is worth to consider the influence of octahedral tilting on the material properties of SrNbO$_3$.

In this study we first address the atomic and electronic structure for SrNbO$_3$ as a function of epitaxial strain using DFT. Since SrTiO$_3$ is a very common substrate for growing perovskite oxides and has been investigated extensively before, we used it as a reference for all calculations. Both compressive and tensile biaxial strain is considered, and the applied strain is always in the (001)-plane of SrTiO$_3$ and SrNbO$_3$. Different octahedral tilts have been investigated with respect to the imposed strain, and the stabilization of the different octahedral tilts with respect to doping was analyzed. Furthermore, we investigate how the excitation spectra and degree of electronic correlations of SrNbO$_3$ evolve as a function of octahedral tilting.
The Seebeck coefficient and the optical loss function have been studied for relevant octahedral tilting modes and angles. We aimed at establishing a complete picture of octahedral tilting in SrNbO$_3$ for various strains.

\section{Computational Methods}
Density functional theory (DFT)~\cite{Kohn1965} was used to investigate the atomic and electronic structures of
SrNbO$_3$ (and SrTiO$_3$). Our DFT calculations were performed using the exchange-correlation functional PBEsol, and projector augmented wave (PAW) potentials~\cite{Blochl1994, Joubert1999} as implemented in \textit{Vienna Ab initio Simulation Package} (VASP).~\cite{Kresse1996} The PBEsol functional was chosen due to its ability of predicting accurate lattice parameters as pointed out in Refs.\cite{Perdew2008, Aschauer2014}. The total energies were calculated with $\Gamma$-centered k-point meshes $8\times8\times8$ and $4\times4\times4$ for the primitive (5 atom) cell and $2\times2\times2$ (40 atom) supercell, respectively. 
The self-consistent loops were converged below \SI{1e-6}{\electronvolt} and the plane-wave energy cutoff was set to \SI{550}{\electronvolt}. A force tolerance of \SI{0.01}{\electronvolt/\angstrom} was set during the relaxations.

Using the DFT obtained relaxed structures, we performed DFT plus dynamical mean field theory (DMFT) calculations at 200 K for SrNbO$_3$ using the \textit{state-of-the-art} fully charge self-consistent implementation.~\cite{DMFTimple} The DFT part were done within Perdew-Burke-Ernzehof generalized gradient approximation (PBE-GGA),~\cite{pbe} as implemented in Wien2K package.~\cite{wien} The DMFT impurity problem was solved by using continuous
time quantum Monte Carlo (CTQMC) calculations~\cite{ctqmc}, with a
Hubbard $U = 6.0$ eV and Hund’s coupling $J = 0.8$ eV.

The Seebeck coefficient was predicted within the constant relaxation time approximation employing BoltzTrap2.~\cite{Madsen2018} This was done using a charge density calculated with a k-point density corresponding to $15\times15\times15$ for the 5 atom cubic unit cell. Non-self-consistent calculations were performed using a k-point mesh of $50\times50\times50$ for the 5 atom cubic unit cell. The electronic states were then interpolated on a grid 15 times as dense using BoltzTrap2. The temperature was set to \SI{300}{\kelvin}.

The complex dielectric function was calculated using the independent-particle random phase approximation (RPA) as implemented in VASP. This implies that the excitations are assumed to be independent, given by the bare Kohn-Sham band structure, and neglect of the local field effects.~\cite{Gajdos2006} A phenomenological Drude term was added to model the contribution of intraband transitions, see Appendix \ref{app:drude}. The imaginary part was set to \SI{0.3}{\electronvolt}, which creates good agreement with experimental loss functions in Reference \cite{Mirjolet2021}.

\section{Results and discussions}

\subsection{Energy landscape of octahedral tilts}
The lattice parameters for varying degrees of biaxial strain were calculated for SrNbO$_3$. This was done using the high symmetry 5 atom unit cells, i.e. excluding any symmetry breaking octahedral tilting. The relative difference, $(a_\mathrm{DFT} -a_\mathrm{exp})/a_\mathrm{exp}$, was ca. \SI{-0.15}{\percent} for unstrained SrNbO$_3$. Here, $a_\mathrm{DFT}=\SI[round-mode=places,round-precision=4]{4.0181816879}{\angstrom}$ and $a_\mathrm{exp}=\SI[round-mode=places,round-precision=3]{4.023}{\angstrom}$ is the predicted and experimental unstrained lattice parameter, respectively. A table of the relaxed out-of-plane lattice parameters for different biaxial strains between \SI{-3}{\percent} and \SI[retain-explicit-plus]{+3}{\percent} can be found in Appendix \ref{app:latparam}. These lattice parameters were fixed throughout the rest of the study.

Using the calculated lattice parameters, we predicted the energy landscapes of the octahedral tilting for varying degrees of strain.
In the case of unstrained SrNbO$_3$ (and SrTiO$_3$) the following pure octahedral tilts were investigated, (using Glazer's notation~\cite{Glazer1972}) $a^0a^0c^-$, $a^0a^0c^+$, $a^0b^-b^-$, $a^0b^+b^+$, $a^-a^-a^-$, and $a^+a^+a^+$. Combined \textit{in-phase} and \textit{out-of-phase} tilts were also sampled, e.g. $a^+a^-a^+$, $a^-a^-a^+$, and $a^0b^-b^+$. The letters $a$, $b$ and $c$ correspond to the rotation angles around $x$, $y$ and $z$ axis, respectively, and the $+/-$ signs denotes \textit{in-phase} and \textit{out-of-phase} rotation between two adjacent octahedra, see Figure \ref{fig:octahedron}. The superscript 0 denotes that no rotation is performed around that Cartesian axis.
For the strained systems we examined octahedral tilts around both the \textit{in-plane} axes and \textit{out-of-plane} axis, since they are no longer symmetrically equivalent.
These calculations were performed on fixed tilts, i.e., for each structure the energy was only evaluated once (self-consistently) without any update of the atomic positions. In this work, we did not consider tilts with different angles around different axes, i.e., it is assumed that the rotation angles are the same around all axes (except if one or more is zero). The stability of $a^0a^0c^+$ and $a^0a^0c^-$ was confirmed by including a small rotation around the $x$ and the $y$ axis. Minimizing the ionic forces lead to a suppression of these rotations, which suggests that the lowest energies are in fact found for systems with rotation \textit{only} around the \textit{out-of-plane} axis in the case of compressive strain.

\begin{figure*}[ht]
    \centering
    \includegraphics[width=\textwidth]{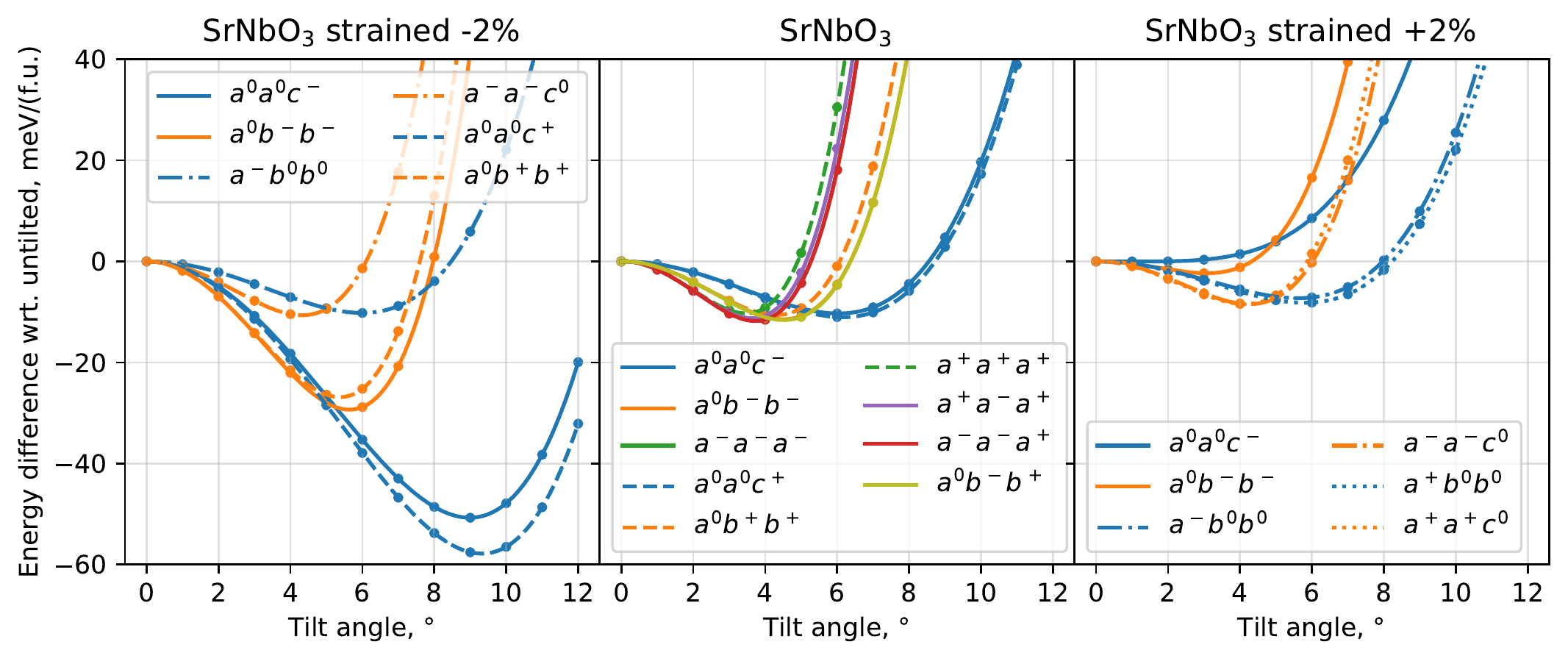}
    \caption{Energy landscapes of octahedral tilting in SrNbO$_3$ under different amounts of biaxial strain. The energy shown is the difference in total internal energy between the tilted structure and untilted structure. Middle plot is the case of no biaxial strain. Compressive strain stabilizes the tilts around the out-of-plane axis, while tensile strain leads to preferred rotations around the in-plane axes. The \textit{in-phase} tilt $a^0a^0c^+$ is found to have the largest energy gain in compressively strained SrNbO$_3$.}
    \label{fig:SNO_tilts_selected}
\end{figure*}

\begin{figure*}[ht]
    \centering
    \includegraphics[width=\textwidth]{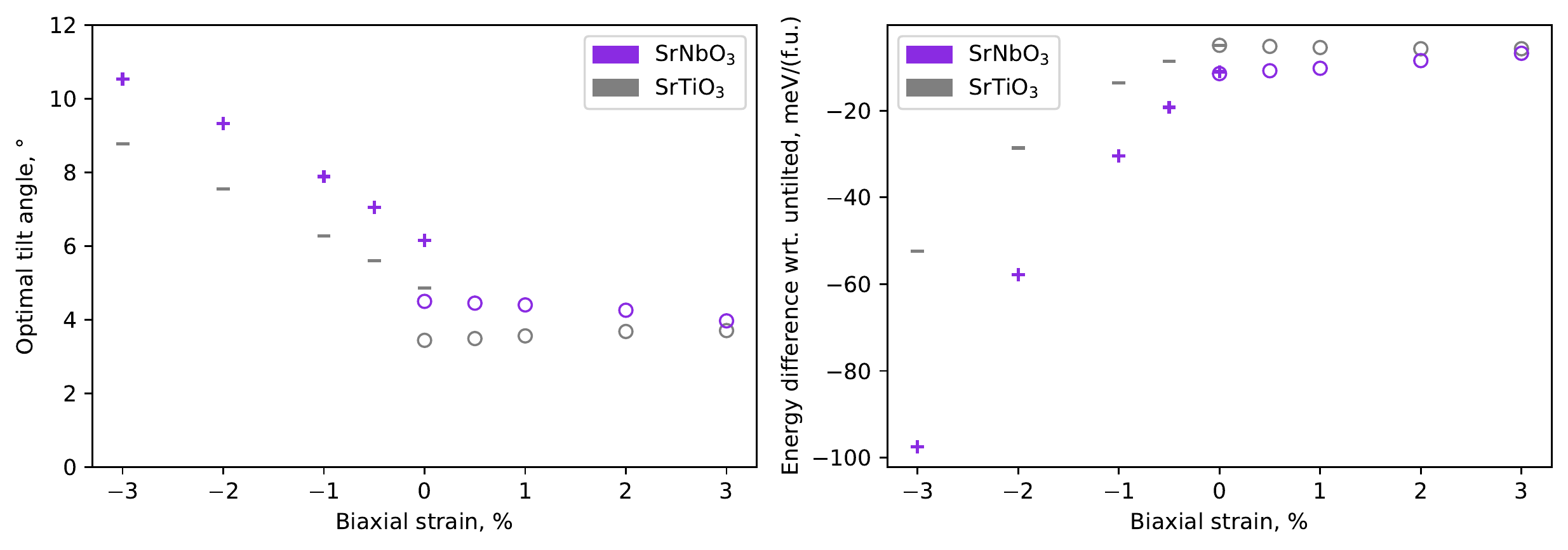}
    \caption{Optimal tilt angles for SrNbO$_3$ and SrTiO$_3$ with varying biaxial strain. The plus and minus marker means that $a^0a^0c^+$ and $a^0a^0c^-$ are lowest in energy, respectively. The circle marker denotes the $a^-a^-c^0$ tilting mode. For \SI{0}{\percent} we include multiple data points, since the energies are very similar and it helps with understanding the limit from compressive to unstrained and tensile to unstrained.}
    \label{fig:optimal_tilts}
\end{figure*}

Figure \ref{fig:SNO_tilts_selected} shows the total internal energy landscapes for different octahedral tilts in SrNbO$_3$ under varying degrees of strain. The reference energy is the total internal energy of the untilted structures, so that the energy goes to zero at zero tilt angle (each strain has its own reference energy). For unstrained SrNbO$_3$, all octahedral tilts show an energy reduction, including the \textit{in-phase} tilting modes. This is not the case for the unstrained SrTiO$_3$ (see  Figure \ref{fig:optimal_tilts} or Appendix \ref{app:alltilts}). This is a notable feature, since the \textit{in-phase} tilting is rare in oxide perovskites.~\cite{Young2016} Furthermore, the energy gain is larger than in SrTiO$_3$ and the optimal tilt angles are also slightly larger. For \SI{-2}{\percent} strained SrNbO$_3$ (SrTiO$_3$) the $a^0a^0c^-$ tilt mode shows a gain of circa \SI{50}{\milli\electronvolt\per\fu} (\SI{30}{\milli\electronvolt\per\fu}) compared to untilted phase and the optimal tilt angle is circa \SI{9}{\degree} (\SI{7.5}{\degree}). The complete data set for SrTiO$_3$ and SrNbO$_3$ with additional strain values can be found in Appendix \ref{app:alltilts}.

By biaxially compressing the oxides (while relaxing the out-of-plane lattice parameter) the B-O bond length in the octahedral structure is contracted in-plane and extended out-of-plane. As a result, the oxygen network is distorted by octahedral tilting. This phenomena can be seen in the left plot in Figure \ref{fig:SNO_tilts_selected}, where the crystals are biaxially strained by \SI{-2}{\percent} (with the out-of-plane lattice parameters relaxed, further details given in Table \ref{tab:latparams} in Appendix \ref{app:latparam}). The key messages are: i) due to the tetragonality, octahedral tilting is preferred around the (longer) out-of-plane axis ii) tilts around one and two axes are no longer degenerate (likely due to competition between B-O and A-O bond lengths, e.g. the A-O bond length is shorter for $a^0b^-b^-$ than $a^0a^0c^-$ and the B-O bond lengths are only slightly longer for $a^0b^-b^-$ than $a^0a^0c^-$) and iii) \textit{in-phase} tilts are energetically favorable for strained SrNbO$_3$, in contrast to SrTiO$_3$ where \textit{out-of-phase} tilting is favorable. The energy gain for \textit{in-phase} tilting relative to \textit{out-of-phase} tilting in SrNbO$_3$ increases with larger biaxial compressive strain, see Appendix \ref{app:alltilts}.
This behaviour is interesting, because perovskites oxides typically show preference for \textit{out-of-phase} tilting~\cite{Young2016}. Furthermore, in a recent study where SrNbO$_3$ thin film was grown on SrTiO$_3$, x-ray diffraction experiments suggests that the $a^0a^0c^-$ tilting mode is stabilized under compressive strain.~\cite{Ok2021}

Biaxial tensile stress elongates the lattice in-plane while the lattice contracts out-of-plane, see Table \ref{tab:latparams}. The resulting energy landscapes with respect to octahedral tilting for \SI{+2}{\percent} biaxially strained SrNbO$_3$ is shown in Figure \ref{fig:SNO_tilts_selected}. The calculated wells are quite shallow compared to the case of compressive strain, instead their depths are similar to the unstrained case. During tensile strain the two in-plane lattice parameters are found to increase and the out-of-plane lattice parameter decreases. Therefore, oxygen octahedral tilting can only alleviate stress in one direction (out-of-plane) by octahedral rotation around the two in-plane axes. In other words, tensile biaxial strain leads to octahedral tilting around the in-plane axes, in contrast with compressive biaxial strain that leads to tilting around the out-of-plane axis, in agreement with investigations of other perovskite oxides.~\cite{Johnson-Wilke2013, Moreau2017} Due to the small energy gains observed here for tensile strained SrNbO$_3$, the possibility of stabilizing octahedral tilting by tensile strain is quite small, especially at elevated temperatures.

The overall optimal tilt modes and their magnitudes are shown in Figure \ref{fig:optimal_tilts}. Here we include SrTiO$_3$ as a reference. It is seen that the octahedral tilting is energetically more stable in the case of SrNbO$_3$ than for the case of SrTiO$_3$ (see also Appendix \ref{app:alltilts}). The optimal tilt angles are ca. \SI{25}{\percent} larger in SrNbO$_3$ than SrTiO$_3$, which is true under compressive strain and small tensile strains. For larger tensile strains the two materials show similar tilting behavior. It is interesting to note that the trends are reversed in tensile strained SrNbO$_3$ and SrTiO$_3$. Larger tensile strain is found to destabilize the tilting (both smaller energy gains and rotation angles) in SrNbO$_3$. The opposite is found in SrTiO$_3$.

In perovskite oxides, \textit{out-of-phase} tilting (or tilts with both \textit{in-phase} and \textit{out-of-phase} components) is the most prevalent tilting mode. Young and Rondinelli~\cite{Young2016} studied bromide and iodide perovskites and linked the octahedral tilt stability to electrostatic interactions, bond valency sums of the A-site ions and charge distribution between the A-site ions and the cations. Here, the A-site ions are not displaced during octahedral tilting, hence the bond valency remains constant when comparing \textit{out-of-phase} ($a^0a^0c^-$) and \textit{in-phase} ($a^0a^0c^+$) tilts. This is because the bond valency is a function of the nearest neighbour distances only.
 
To understand what stabilizes the \textit{in-phase} tilt in SrNbO$_3$ and SrTiO$_3$ we controlled doping by creating holes in SrNbO$_3$ and adding electrons in the case of SrTiO$_3$. We added (or removed) charge in steps of 0.25 electrons per formula unit up to 1 electron per formula unit. Additional electrons can increase the energy gain by octahedral tilting in SrTiO$_3$, as indicated in Reference \cite{Uchida2003}. Hence, octahedral tilting is stabilized with the addition of electrons in SrTiO$_3$. Uchida \textit{et al.}~\cite{Uchida2003} attribute the stabilization of the tilting to the increase in size of the Ti ion, hence decreasing the Goldschmidt factor. The Goldschmidt tolerance factor, $t = (r_\mathrm{A} + r_\mathrm{O})/(\sqrt{2} (r_\mathrm{B} + r_\mathrm{O}))$ where $r_i$ denotes the ionic radius of ion $i$, indicates the stability of octahedral distortions in perovskite oxides~\cite{Goldschmidt1926}. In the following we will focus only on \SI{-2}{\percent} strained system since it is a relevant strain observed in thin films. The resulting energy landscapes are shown in Figure \ref{fig:energy_angle_charge} in Appendix \ref{app:tiltcharge}. By analysing both the $a^0a^0c^-$ and the $a^0a^0c^+$ tilting modes it is evident that including electrons reduces the energy difference between the two tilt modes. To the best of our knowledge, this has not been considered in earlier works. On the contrary, introducing holes to SrNbO$_3$ is found to destabilize the two tilt modes. Here, as in the case of SrTiO$_3$, changing the number of electrons (by addition of holes) also changes the energy difference between the two tilt modes. Interestingly, there is a critical point at which $a^0a^0c^-$ is stabilized over $a^0a^0c^+$ in SrNbO$_3$ (see Figure \ref{fig:energy_angle_charge}). Adding 0.5 holes per formula unit, i.e. 0.5 holes per Nb ion, to SrNbO$_3$ makes the two tilt modes almost degenerate with a slight preference for $a^0a^0c^-$ over $a^0a^0c^+$.

This suggests that the stability of the \textit{out-of-phase} ($a^0a^0c^-$) and \textit{in-phase} ($a^0a^0c^+$) tilting modes is connected to the number of electrons and the size of the A ion (e.g. Nb and Ti).
Further investigations by doping with different ion sizes could benefit our understanding of the stability of the various tilting modes in more complex scenarios. These results also suggest that the so-called rigid-band assumption should be used with caution for perovskites. In other words, it is possible that electron doping affects the octahedral tilting stability, hence the assumption that doping can be captured by a simple shift of Fermi level is questionable. In summary, these predictions suggest that it is possible to tune the stability of octahedral tilting by doping and/or gating.

\subsection{Effects of tilt and strain on band structure and spectral function}
We then moved over to the electronic structure of SrNbO$_3$. The electronic band structure of SrNbO$_3$ was studied with different octahedral tiltings. Here, we focus on $a^0a^0c^-$ and $a^0a^0c^+$ since they are both stabilized under compressive biaxial strain. The reference in these cases was the untilted ($a^0a^0a^0$) SrNbO$_3$.

\begin{figure*}[ht]
    \centering
    \includegraphics[width=\textwidth]{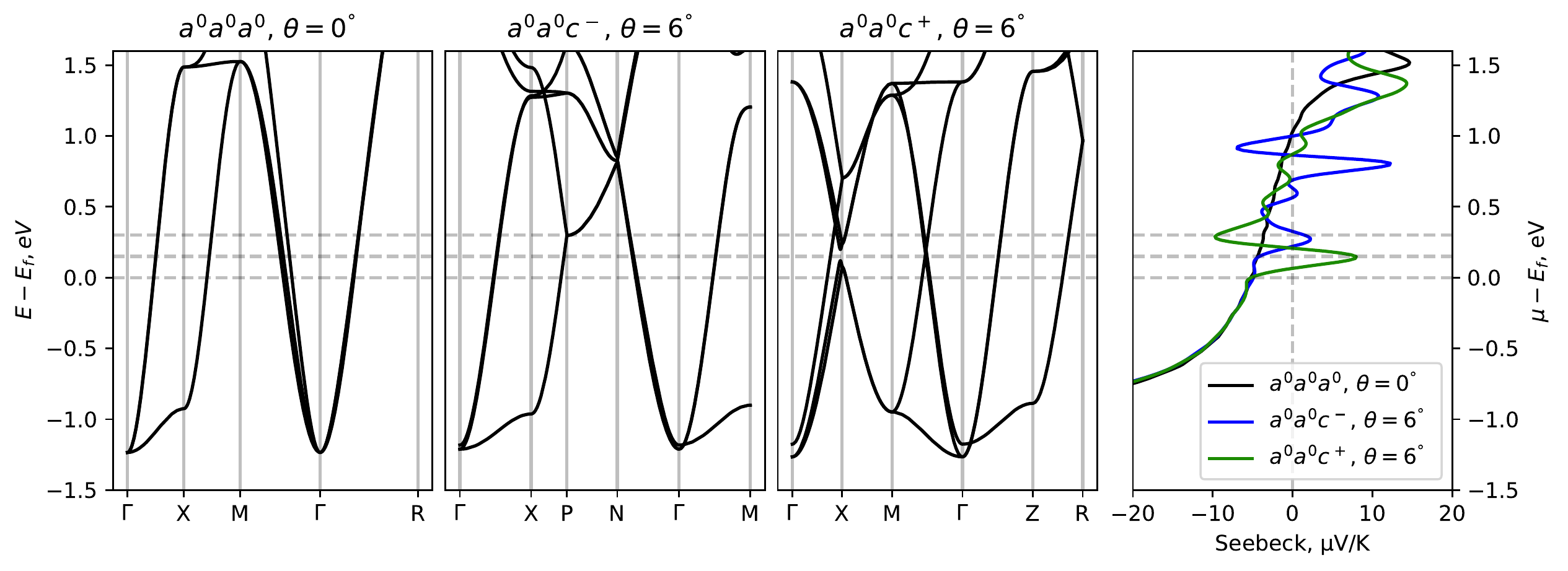}
    \caption{Conduction bands of unstrained SrNbO$_3$ with $a^0a^0a^0$, $a^0a^0c^-$ and $a^0a^0c^+$ tilt modes. The tilt angle, $\theta$, is set to \SI{6}{\degree}. To the right, the in-plane Seebeck coefficient as a function of chemical potential at \SI{300}{\kelvin} is shown. The peaks in the Seebeck coefficient align with the (avoided) band crossings at $P$ for $a^0a^0c^-$ and at $X$ as well as along $M - \Gamma$ for $a^0a^0c^+$. The $\Gamma$-point $t_{2g}$ split is sensitive to the tilting mode. For $a^0a^0a^0$, the $t_{2g}$ states are degenerate at $\Gamma$. The degeneracy is lifted for $a^0a^0c^-$ and $a^0a^0c^+$. The Brillouin zones, with sampled high symmetry points, for the different tilting modes can be found in Appendix \ref{app:BZ}.}
    \label{fig:bs-and-seebeck}
\end{figure*}

In Figure \ref{fig:bs-and-seebeck} the conduction bands for unstrained SrNbO$_3$ is shown for different octahedral tilts. The band structures are calculated along high symmetry points of the first Brillouin zone for the different crystals (see Appendix \ref{app:BZ} for Brillouin zones). Here, the rotation angle is set to \SI{6}{\degree} which is close to the minima for $a^0a^0c^-$ and $a^0a^0c^+$ in the unstrained case. The untilted system shows the heavy and light bands which originate from the $t_{2g}$-like niobium orbitals. Since the tilts requires repetitions of the minimal unit cell, there are additional bands in $a^0a^0c^-$ and $a^0a^0c^+$. The \textit{out-of-phase} tilt $a^0a^0c^-$ shows a semi-Dirac point at $P$, in agreement with Reference \cite{Ok2021}. This tilt mode has been further investigated theoretically illustrating the tunability of the Berry phase and the anomalous Hall coefficient.~\cite{Mohanta2021} A similar dispersion is found at the $X$-point in $a^0a^0c^+$, albeit at a substantially higher energy. Both tilt modes create $t_{2g}$ splitting at the $\Gamma$-point, however for $a^0a^0c^-$ with \SI{6}{\degree} rotation angle the splitting is quite small. The \textit{in-phase} tilt $a^0a^0c^+$ shows a gap opening at the $X$-point slightly above the Fermi level. The band velocities are low at the bottom of the conduction bands and higher near the Fermi level, as can be seen from the curvature of the bands ($\hbar v_{\mathbf{k}}  = \partial_\mathbf{k} \epsilon_\mathbf{k}$). Moreover, the bands near the Dirac points show both mobile and slow carriers, as indicated by the curvature of the bands. As example, the states from $P$ to $X$ show high velocities, while the opposite is true for the states from $P$ towards $N$ in the case of $a^0a^0c^-$.

The qualitative picture of the band structure is rather similar for compressively strained SrNbO$_3$ (see Appendix \ref{app:allbs}). There are however a few noteworthy changes. Straining introduces $t_{2g}$ splitting in the untilted system, due to a symmetry lowering from cubic to tetragonal. Inclusion of octahedral tilting further splits the $t_{2g}$ states. For \SI{-2}{\percent} strained SrNbO$_3$, the split is ca. \SI{0.15}{\electronvolt} and \SI{0.3}{\electronvolt}, for $a^0a^0c^-$ and $a^0a^0c^+$, respectively, both with \SI{9}{\degree} tilt angle. By examining the splitting for different strain-tilt combinations, it is observed that the large $t_{2g}$ splitting in strained SrNbO$_3$ originates from both the strain alone and from the increase in tilt angle with strain. The $t_{2g}$ splitting at the $\Gamma$ point can be found in Appendix \ref{app:t2g}. As an example, the $t_{2g}$ splitting in \SI{-1}{\percent} strained $a^0a^0a^0$ is ca. \SI{0.05}{\electronvolt} and is of similar magnitude to the splitting in unstrained $a^0a^0c^+$ with \SI{6}{\degree} tilt angle. Furthermore, the Dirac point at $P$ is shifted closer to the Fermi level for $a^0a^0c^-$ with compressive strain. While the gapping at $X$ is larger in the case of \SI{-2}{\percent} strained $a^0a^0c^+$ with \SI{9}{\degree} rotation angle, than unstrained $a^0a^0c^+$ with \SI{6}{\degree}. The increase in the gap at $X$ is dominated by the enhancement of the tilt angle with strain, i.e., the gap is not so sensitive to the strain alone. We speculate that it in fact may be the $t_{2g}$ splitting which stabilize the $a^0a^0c^+$ tilt mode by decreasing the lowest conduction band energies at the $\Gamma$-point and creating a gap at $X$, compared to the other modes, see Figure \ref{fig:bs-and-seebeck}.

\begin{figure*}[ht]
    \centering
    \includegraphics[scale=0.4]{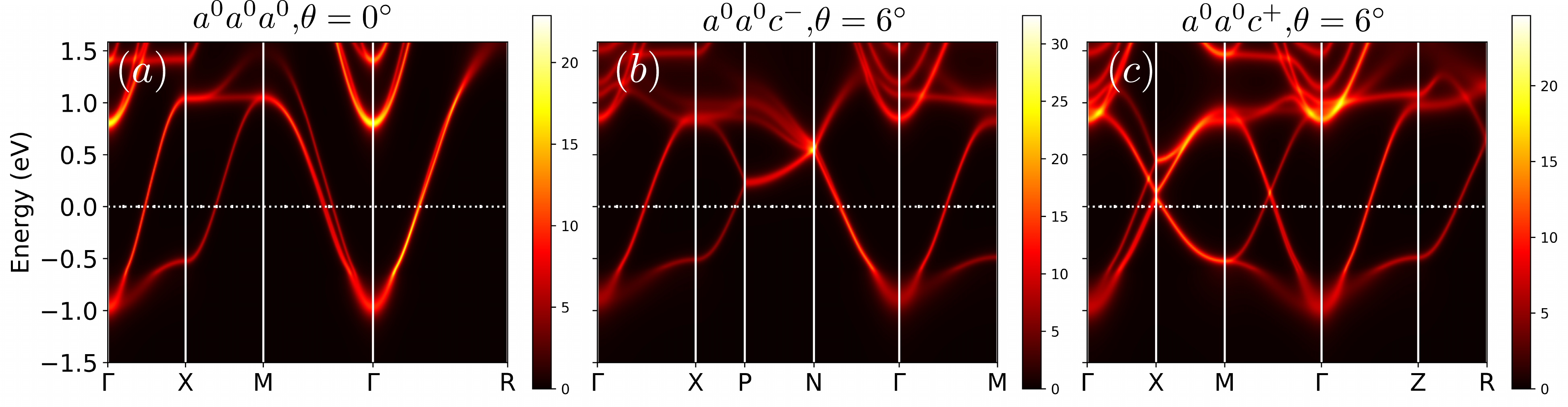}
    \caption{DFT+DMFT calculated spectral functions at 200 K of unstrained SrNbO$_3$ with $a^0a^0a^0$ (a), $a^0a^0c^-$ (b) and $a^0a^0c^+$ (c) tilt modes. }
    \label{fig:Awk_SrNbO3}
\end{figure*}

In Figure \ref{fig:Awk_SrNbO3}, we present the DFT+DMFT obtained spectral functions. As can be noticed, the dynamical correlations downshift the Dirac like points toward the Fermi level. 
The $P$ and $N$ crossing points exhibited in  the $a^0a^0c^-$ structure are ca. \SI{0.25}{\electronvolt} and \SI{0.55}{\electronvolt}, above the Fermi energy. Compared to \SI{0.30}{\electronvolt} and \SI{0.75}{\electronvolt}, respectively, in the DFT band structure. Furthermore, we observe an additional crossing point at \SI{0.27}{\electronvolt} along $\Gamma-X$ in $a^0a^0c^+$ structure. The Dirac point between $M-\Gamma$ appears around \SI{0.15}{\electronvolt}. 

Overall, these findings indicate that the interplay of lattice distortions and electronic correlations is a key factor for the topology properties of strained SrNbO$_3$ thin films. This analysis also illustrates the tunability of the electronic states in SrNbO$_3$ with biaxial compressive strain. The tunability has two components, there is an effect from straining alone, but also an effect from the tilt that is induced by the strain. Furthermore, since the stability of the octahedral tilting modes seems to be connected to the number of electrons, it could also act as a turning knob for the atomic structure and hence electronic structure additional to a shift of Fermi level.

\subsection{Seebeck coefficient and optical properties}
\begin{figure*}[ht]
    \centering
    \includegraphics[width=\textwidth]{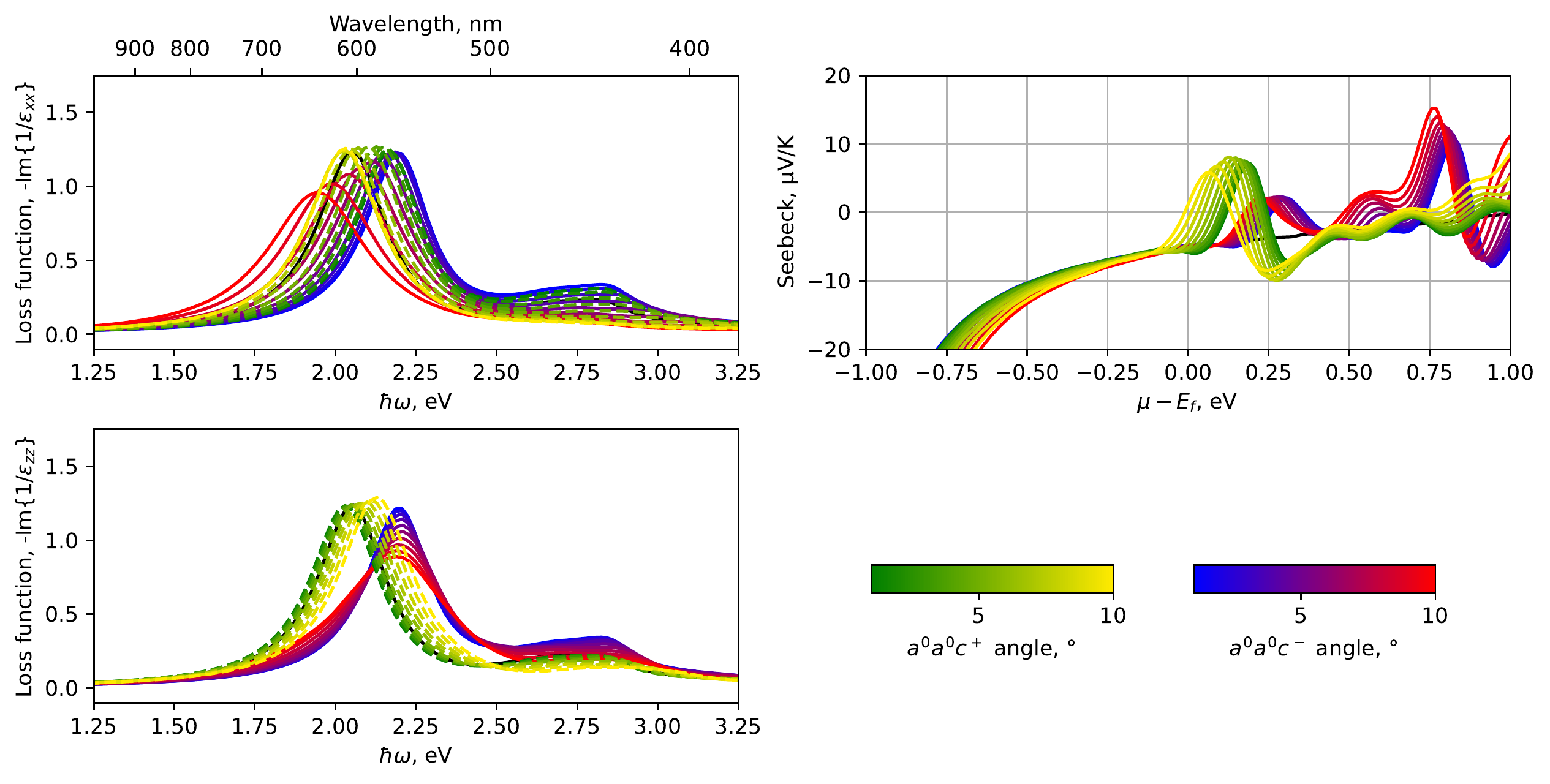}
    \caption{Influence of octahedral tilting on the optical loss function (top left and bottom left) and on the room temperature in-plane Seebeck coefficient (top right) in unstrained SrNbO$_3$. The black lines correspond to the response of untilted SrNbO$_3$. In the upper left (lower left) figure the $xx$ ($zz$) component of the optical loss function is shown. For visibility, the $a^0a^0c^+$ tilts are drawn with a dashed line. A systematic redshift with tilt angle is observed in the loss peak of the $xx$ component, while the $zz$ component shows a blueshift with tilt angle of $a^0a^0c^+$ and an almost constant peak position for $a^0a^0c^-$. Due to the (avoided) band crossings related to octahedral tilting the Seebeck coefficient shows sign changes over small range of chemical potentials.}
    \label{fig:seebeck_and_loss}
\end{figure*}

We further investigated the influence of the octahedral tilting on the Seebeck coefficient. The Seebeck coefficient was predicted since within the constant relaxation time approximation the Seebeck coefficient is independent of the scattering rate~\cite{Madsen2018}, making it suitable as a probe of the electronic structure. Hence, changes in the band structure could have an effect on the Seebeck coefficient which could act as a fingerprint for the octahedral tilting. In Figure \ref{fig:seebeck_and_loss}, the in-plane (i.e. $xx=yy$ component) Seebeck coefficient is shown for the two tilts as a function of octahedral rotation angle. The results are shown for unstrained SrNbO$_3$ at a temperature of \SI{300}{\kelvin}. The qualitative trends are same for \SI{-2}{\percent} strained SrNbO$_3$. Below ca. \SI{-100}{\milli\electronvolt} all configurations show a negative Seebeck coefficient lower than \SI{-5}{\micro\volt\per\kelvin}, corresponding to n-type transport. Typical Seebeck coefficients for metals or heavily doped semiconductors is $\pm$ \SI{10}{\micro\volt\per\kelvin}.~\cite{Rowe2018} Interestingly, near the Fermi level and slightly above, the Seebeck coefficient varies and can even become positive, when octahedral tilting is included. Furthermore, this behaviour is different for $a^0a^0c^-$ and $a^0a^0c^+$. The positive values occur at lower energies for $a^0a^0c^+$ than $a^0a^0c^-$. In the case of $a^0a^0c^+$ there is also a sharp decrease in the Seebeck coefficient after the increase with higher chemical potential. These features can be connected to the band crossings in Figure \ref{fig:bs-and-seebeck}. The semi-Dirac point at $P$ coincides with the first Seebeck peak in the case of $a^0a^0c^-$. Similarly, in the case of $a^0a^0c^+$ the first peak coincides with the gapped bands at $X$ and the cluster of bands along $M - \Gamma$. The sign of the Seebeck coefficient is sensitive to the character of the dispersion, such as non-parabolic features and band crossings, as indicated in Reference \cite{Wei2009} for graphene. This highlights the difference between the effects of strain and octahedral tilting, since strain alone does not create new features such as band crossings and gapping near the Fermi level. The Seebeck values are, of course, small, i.e. not relevant for thermoelectric generator purposes, but could be valuable as fingerprints of octahedral tilting.
Measurements of the Seebeck coefficient, with varying gate voltage, of SrNbO$_3$ could be used as an indirect probe of the electronic structure. Abrupt sign changes, with respect to chemical potential, could be used as a fingerprint of octahedral tilting and an indication of change in the topology of the bands, compared to the bands of untilted SrNbO$_3$ that shows a much flatter change in Seebeck with respect to chemical potential.

We then examined the optical properties of SrNbO$_3$ for different octahedral tilting by calculating the optical loss function. The optical loss function, which describes the interaction between electromagnetic fields and matter, is an indirect probe of the electronic states. The optical loss function can be written as $LF(\omega)=-\Im\{\epsilon(q\rightarrow0, \omega)^{-1}\}$, where the optical limit $q\rightarrow0$ of the dielectric function, $\epsilon(\omega)$, results in no momentum transfer between the applied field and the electrons.
The loss function exhibits a peak when the real part of the dielectric function changes sign (a slight energy shift can occur due to the frequency dependent imaginary part). Therefore, the loss function peak position is a good indication of the plasma frequency defined by $\Re\{\epsilon(\omega_p)\} = 0$.
The optical loss functions given by the independent-particle random phase approximation are shown in Figure \ref{fig:seebeck_and_loss} for unstrained SrNbO$_3$. Both $xx$ and $zz$ components of the loss functions are shown. A clear loss peak is observed in the visible regime for each tilt configuration, as in agreement with References \cite{Mirjolet2021,Park2020}. Therefore, the plasma frequencies are in the visible regime. Increasing the rotation angle in $a^0a^0c^-$ decreases the peak height for the $xx$ and $zz$ components. Furthermore, there is a redshift of the peak (and hence plasma frequency) of the $xx$ component, while this is not true for the $zz$ component. Contrary to this, the peak height seems not to be sensitive to variations of the $a^0a^0c^+$ rotation angle. There is also a redshift of the peak and plasma frequency for $a^0a^0c^+$ with larger rotation angles when the $xx$ component is considered. The opposite is true in the $zz$ component, i.e. there is a blueshift of the plasma frequency with respect to larger rotation angles for $a^0a^0c^+$.
For smaller octahedral rotations there is also a secondary peak at circa \SI{2.75}{\electronvolt}. This peak has been connected to the excitations from $t_{2g}$ to $e_{g}$ bands.~\cite{Park2020} This peak is weakened with larger rotation angle, especially for the $xx$ component. The peaks vanish for large rotations for both $a^0a^0c^-$ and $a^0a^0c^+$. It has been noted that the small peak is absent in experiments, while it is present in DFT. This fits well with our predictions, since octahedral tilting breaks the octahedral crystal field, i.e., the $t_{2g}$ and $e_{g}$ states are altered by octahedral tilting. One should, however, further analyse the transmission matrix elements between these states, with various tilting, for a complete picture. To our knowledge, octahedral tilting was not included in the DFT analysis by Park and co-workers.~\cite{Park2020} It is possible that the missing secondary peak in spectroscopic ellipsometry at room temperature is due to thermally fluctuating oxygen ions in the SrNbO$_3$ rather than stabilization of a tilting mode. Temperature dependent spectroscopic ellipsometry could be valuable for understanding the stabilization of the octahedral tilting modes, given the strong signals due to tilting predicted here and the energy scale involved, which is comparable to the thermal energy.

The independent particle assumption used in the calculation of the dielectric function is relatively crude, but it has been shown to qualitatively reproduce the experimental plasmon peak in SrNbO$_3$ when intraband transitions are included.~\cite{Zhu2018} We modelled intraband transitions by using a phenomelogical Drude term (see Appendix \ref{app:drude}). Although these calculations involve serious approximations, they indicate how some typical experimental quantities can give a fingerprint of the underlying octahedral tilting.

These observations call for further ab initio modelling of SrNbO$_3$ (and other perovskites). Not only by assuming highly symmetric unit cells, but also allowing for deviations from highly symmetric cells by the use of supercells.~\cite{Zhao2021} Further investigations of the stabilization of octahedral tilting with respect to strain under finite temperatures will be highly relevant for application purposes. Our results show that octahedral tilting can have a significant effect on thermoelectric and optical properties.
The Seebeck coefficient exhibits sign changes due to the new band features appearing from the octahedral tilting. Moreover, the optical loss function peak position and plasma frequency is tuneable by octahedral tilting over a range of ca. \SI{100}{\nano\metre} in the visible regime. Therefore, octahedral tilting should be considered when analyzing strained perovskite oxides like SrNbO$_3$.

\section{Conclusions}
A first principles investigation of the atomic and electronic structure of SrNbO$_3$ under biaxial strain has been performed with emphasis on the relation between octahedral tilting and strain. This was done by using supercells to allow for the symmetry breaking octahedral tilting, while the biaxial strain was applied in the (001)-plane. By methodically sampling different tilt modes and magnitudes, the optimal octahedral tilting could be found for the different strain conditions. Compressive (tensile) strain was found to induce octahedral tilting around the out-of-plane (in-plane) axis. Interestingly, the \textit{in-phase} tilt $a^0a^0c^+$ was energetically favorable for SrNbO$_3$ under compressive strain. This is in contrast to $a^0a^0c^-$, which has been reported in a recent experimental study of SrNbO$_3$.~\cite{Ok2021} The \textit{in-phase} tilt, $a^0a^0c^+$, shows larger $t_{2g}$ splitting compared to the \textit{out-of-phase} tilt $a^0a^0c^-$. The electronic dispersion is also gapped slightly above the Fermi level at the $X$-point, in the case of $a^0a^0c^+$. DFT+DMFT calculations show that the (avoided) band crossings due to octahedral tilting are shifted towards the Fermi level by correlations, hence making these points easier to reach experimentally. Furthermore, the Seebeck coefficient shows positive values slightly above the Fermi level for $a^0a^0c^-$ and $a^0a^0c^+$, as compared with the untilted SrNbO$_3$ that only displays negative values. Interestingly, the tilts are distinguishable due to the difference in the required chemical potential needed to observe positive values of the Seebeck coefficient.
The peak in the optical loss function shows tunability with respect to octahedral tilting. Furthermore, the trends in peak position (and hence plasma frequency) and heights are different for the two tilts. We also show that the small feature around \SI{2.75}{\electronvolt} is sensitive to octahedral tilting, which could explain the absence of this peak in experiments, see Reference \cite{Park2020}.\newline
\indent With the recent interest in SrNbO$_3$ we hope to give guidance for how the octahedral tilting behaves in the material but also suggest that other tilting modes, than $a^0a^0c^-$, should be considered and investigated further in perovskite oxides. The pronounced features in thermoelectric and optical properties with respect to octahedral tilting could be used as indication of tilting in SrNbO$_3$, and of the changes in the electronic structure therein.

\begin{acknowledgments}
\noindent This work has been supported by Independent Research Fund Denmark grant 8048-00088B and Innovation Fund Denmark grant 1045-00029B. This project has received funding from the European Union’s Horizon 2020 research and innovation programme under the Marie Skłodowska-Curie grant agreement No 884104 (PSI-FELLOW-III-3i).
\end{acknowledgments}

\clearpage
\appendix
\section{Lattice parameters for 5 atom cell}\label{app:latparam}
The equilibrium lattice parameters were calculated for SrNbO$_3$ and SrTiO$_3$ with varying degrees of biaxial strain. This was done using the high symmetry 5 atom unit cells. The objective is to find lattice parameters that can be used for all the following DFT calculations. For biaxial strain, the lattice parameters are given by $a=b= a_0(\epsilon + 1)$, where $a_0=b_0$ is the relaxed unstrained lattice parameter and $\epsilon$ is the biaxial strain. The results are shown in Table \ref{tab:latparams}.

\begin{table}[ht]
\centering
\caption{Predicted lattice parameters using PBEsol in units of Ångström. In the strained cases the relaxed out-of-plane parameter is presented.}
\label{tab:latparams}
\begin{tabular}{|c|c|c|}
\hline
Biaxial strain, \% & SrNbO$_3$ & SrTiO$_3$ \\ \hline
-3.0 & \num[round-mode=places,round-precision=3]{4.0949573125586065} & \num[round-mode=places,round-precision=3]{3.973956732881168} \\ \hline
-2.0 & \num[round-mode=places,round-precision=3]{4.07309864745539} & \num[round-mode=places,round-precision=3]{3.9473547231683717} \\ \hline
-1.0 & \num[round-mode=places,round-precision=3]{4.046744052128023} & \num[round-mode=places,round-precision=3]{3.9221373280904586} \\ \hline
-0.5 & \num[round-mode=places,round-precision=3]{4.032884064768215} & \num[round-mode=places,round-precision=3]{3.9100144131655052} \\ \hline
0.0 & \num[round-mode=places,round-precision=3]{4.0181635649673586} & \num[round-mode=places,round-precision=3]{3.8957640367923894} \\ \hline
+0.5 & \num[round-mode=places,round-precision=3]{4.0074337621815985} & \num[round-mode=places,round-precision=3]{3.8867862858083226}\\ \hline
+1.0 & \num[round-mode=places,round-precision=3]{3.996853584923392} & \num[round-mode=places,round-precision=3]{3.8757684492792657}\\ \hline
+2.0 & \num[round-mode=places,round-precision=3]{3.9789486620054464} & \num[round-mode=places,round-precision=3]{3.854964621077708} \\ \hline
+3.0 & \num[round-mode=places,round-precision=3]{3.96369087620064} & \num[round-mode=places,round-precision=3]{3.8355513122490104} \\ \hline
\end{tabular}
\end{table}

As expected, the lattice parameter of SrNbO$_3$ is substantially larger than that of SrTiO$_3$.~\cite{Oka2015} Our predicted unstrained lattice parameters were \SI[round-mode=places,round-precision=4]{3.8957640367923894}{\angstrom} and \SI[round-mode=places,round-precision=4]{4.0181816879}{\angstrom}, which is in excellent agreement with the experimental values \SI[round-mode=places,round-precision=3]{3.905}{\angstrom} and \SI[round-mode=places,round-precision=3]{4.023}{\angstrom}, for the titanate and niobate, respectively.~\cite{Oka2015} The difference between our predicted unstrained lattice parameter and the literature value is \SI[round-mode=places,round-precision=3]{-0.0092359632076106}{\angstrom} and \SI[round-mode=places,round-precision=3]{-0.0048183121}{\angstrom}, for SrTiO$_3$ and SrNbO$_3$ respectively. In the following the lattice parameters presented in Table \ref{tab:latparams} are kept fixed for a given strain value.

\section{Drude term for intraband transitions}\label{app:drude}
A phenomelogical Drude intraband term was included in the calculation of the dielectric function:
\begin{equation}
    \epsilon_\mathrm{intra}(\omega) = \epsilon_\mathrm{intra}^{(1)}(\omega) + i\epsilon_\mathrm{intra}^{(2)}(\omega),
\end{equation}
where the real and imaginary parts are given by:
\begin{align}
    \epsilon_\mathrm{intra}^{(1)}(\omega) &= 1 - \frac{\omega_{p,\mathrm{intra}}^2}{\omega^2 + \gamma^2}\\
    \epsilon_\mathrm{intra}^{(2)}(\omega) &= \frac{\gamma\omega_{p, \mathrm{intra}}^2}{\omega^3 + \omega\gamma^2} 
\end{align}
Here $\omega_{p,\mathrm{intra}}$ is the intraband plasma frequency~\cite{Harl2007} and $\gamma = \SI{0.3}{\electronvolt}$ is the inverse lifetime chosen in this work to reproduce the broadening in Reference \cite{Mirjolet2021}.

\newpage
\section{First Brillouin zone for different tilts}\label{app:BZ}
\begin{figure}[ht!]
    \centering
    \includegraphics[width = 0.3\textwidth]{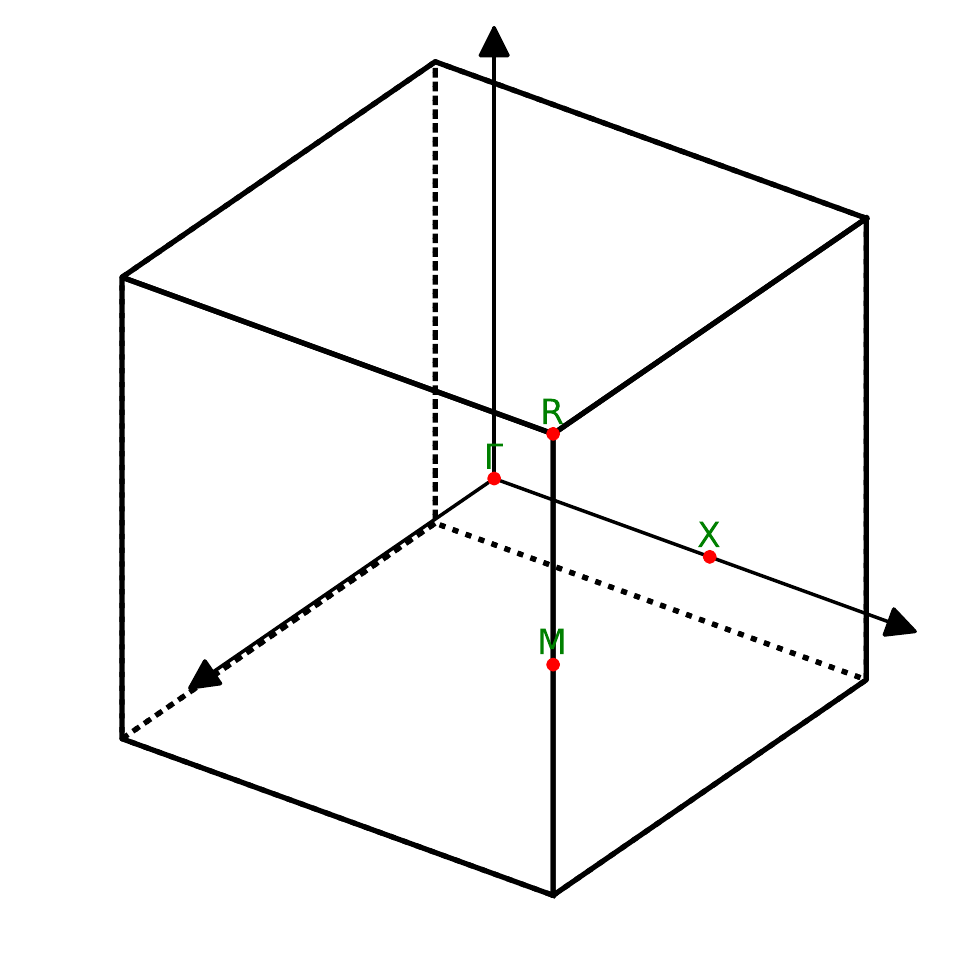}
    \caption{BZ for $a^0a^0a^0$}
    \label{fig:bza0a0a0}
\end{figure}

\begin{figure}[ht!]
    \centering
    \includegraphics[width = 0.3\textwidth]{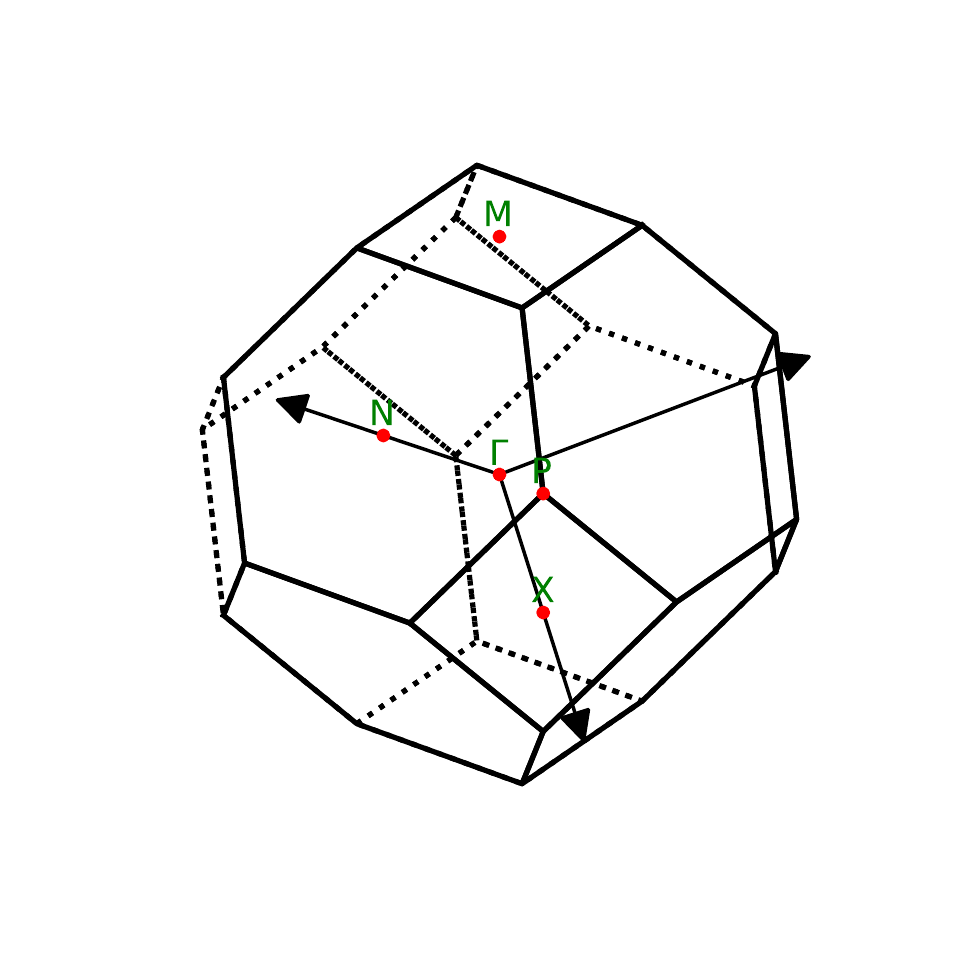}
    \caption{BZ for $a^0a^0c^-$}
    \label{fig:bza0a0c-}
\end{figure}

\begin{figure}[ht!]
    \centering
    \includegraphics[width = 0.3\textwidth]{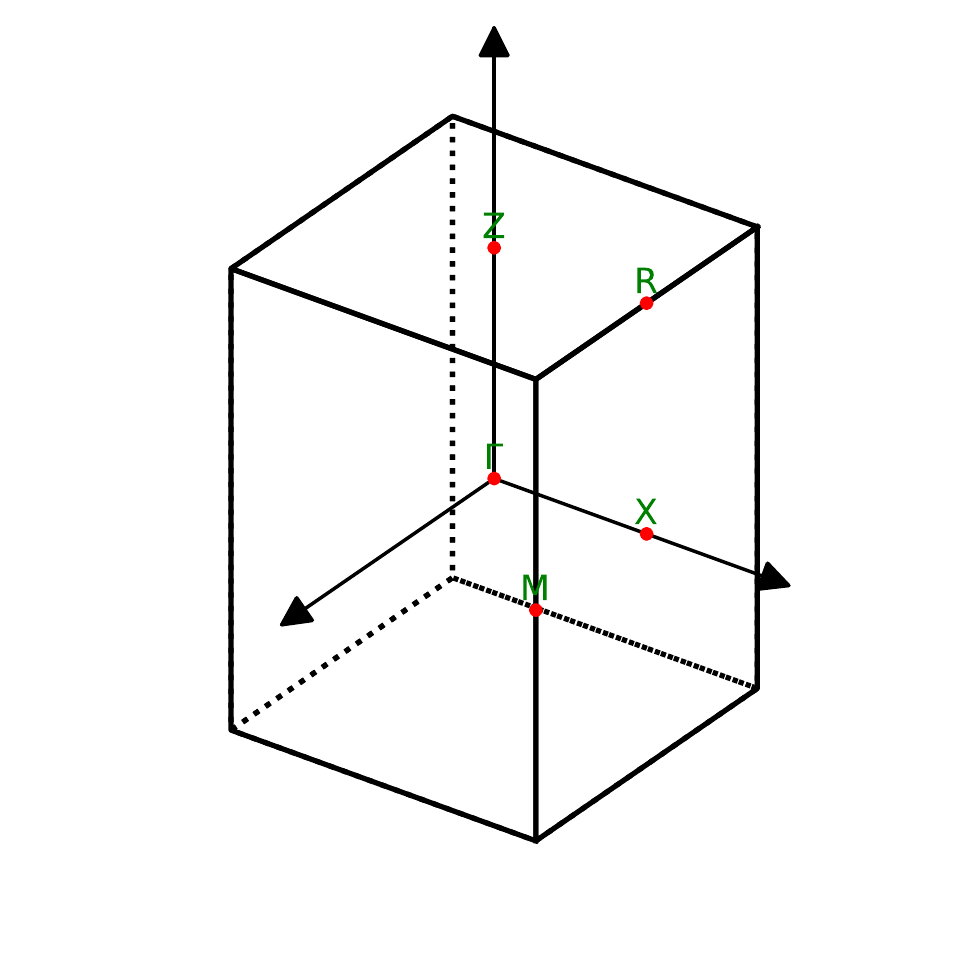}
    \caption{BZ for $a^0a^0c^+$}
    \label{fig:bza0a0c+}
\end{figure}

In Figures \ref{fig:bza0a0a0}, \ref{fig:bza0a0c-} and \ref{fig:bza0a0c+} the first Brillouin zones of $a^0a^0a^0$, $a^0a^0c^-$ and $a^0a^0c^+$ are shown. Note that the coordinate systems for the tilted structures are rotated \SI{45}{\degree} around the $z$-axis of the untilted structure.

\onecolumngrid
\section{Stability of tilts with charge}\label{app:tiltcharge}
\begin{figure*}[ht!]
    \centering
    \includegraphics[width=\textwidth]{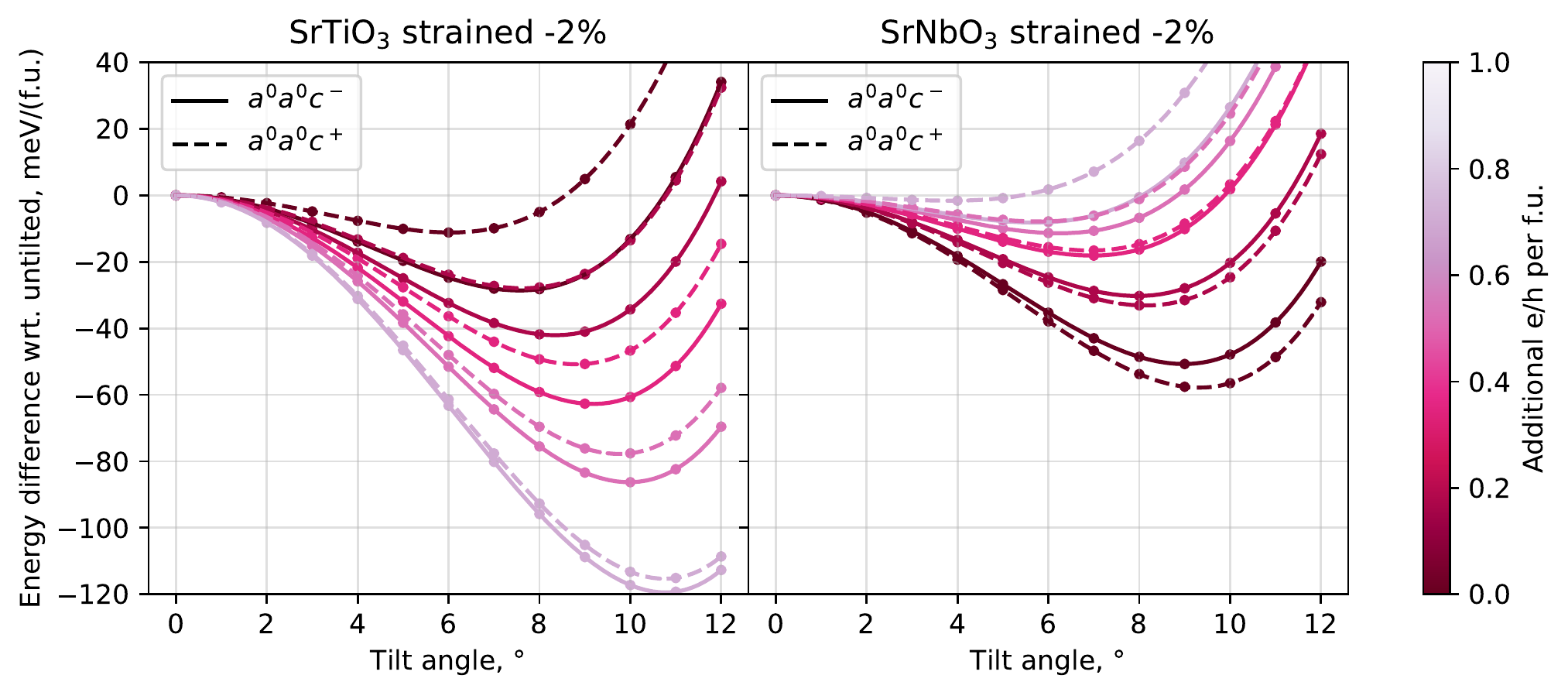}
    \caption{Effect of electron (hole) doping on tilting stability of SrTiO$_3$ (SrNbO$_3$) under \SI{-2}{\percent} biaxial strain. SrTiO$_3$ to the left and SrNbO$_3$ to the right. The darkest purple lines correspond to undoped systems and lighter lines correspond to more electrons (or holes). Adding electrons to SrTiO$_3$ causes the two tilt modes to stabilize. Furthermore, the addition of electron reduces the difference in energy between \textit{in-phase} and \textit{out-of-phase} tilting. SrNbO$_3$ shows destabilization of tilts with addition of holes. For SrNbO$_3$ there is a critical point at which the optimal tilt transitions from \textit{in-phase} to \textit{out-of-phase}.}
    \label{fig:energy_angle_charge}
\end{figure*}

\section{$t_{2g}$ splitting map}\label{app:t2g}
\begin{figure*}[ht]
    \centering
    \includegraphics[width = \textwidth]{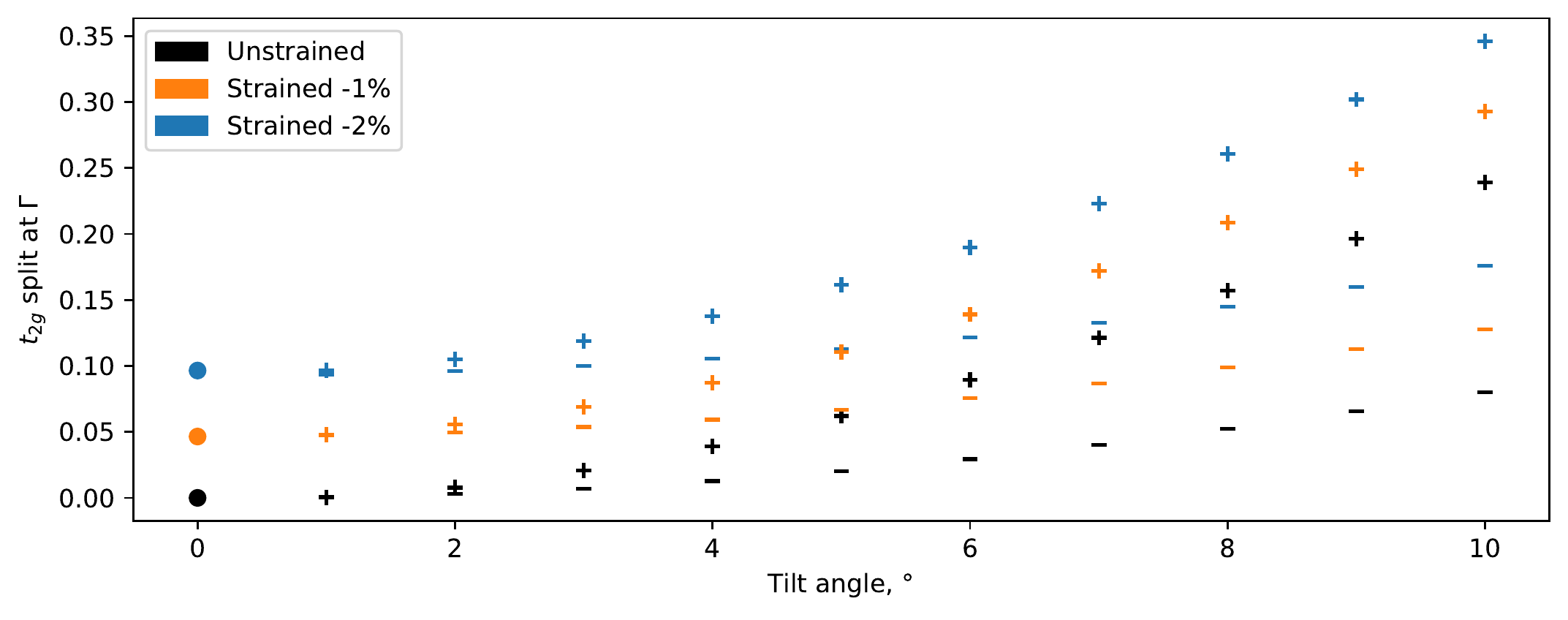}
    \caption{Splitting between the $t_{2g}$ orbitals at $\Gamma$. Circles denote untilted, plus signs denote $a^0a^0c^+$ and minus signs denote $a^0a^0c^-$.}
    \label{fig:t2g_appendix}
\end{figure*}

\section{All octahedral tilts}\label{app:alltilts}
\begin{figure*}[ht!]
    \centering
    \includegraphics[trim={0 36cm 0 0},clip,height=0.85\textheight]{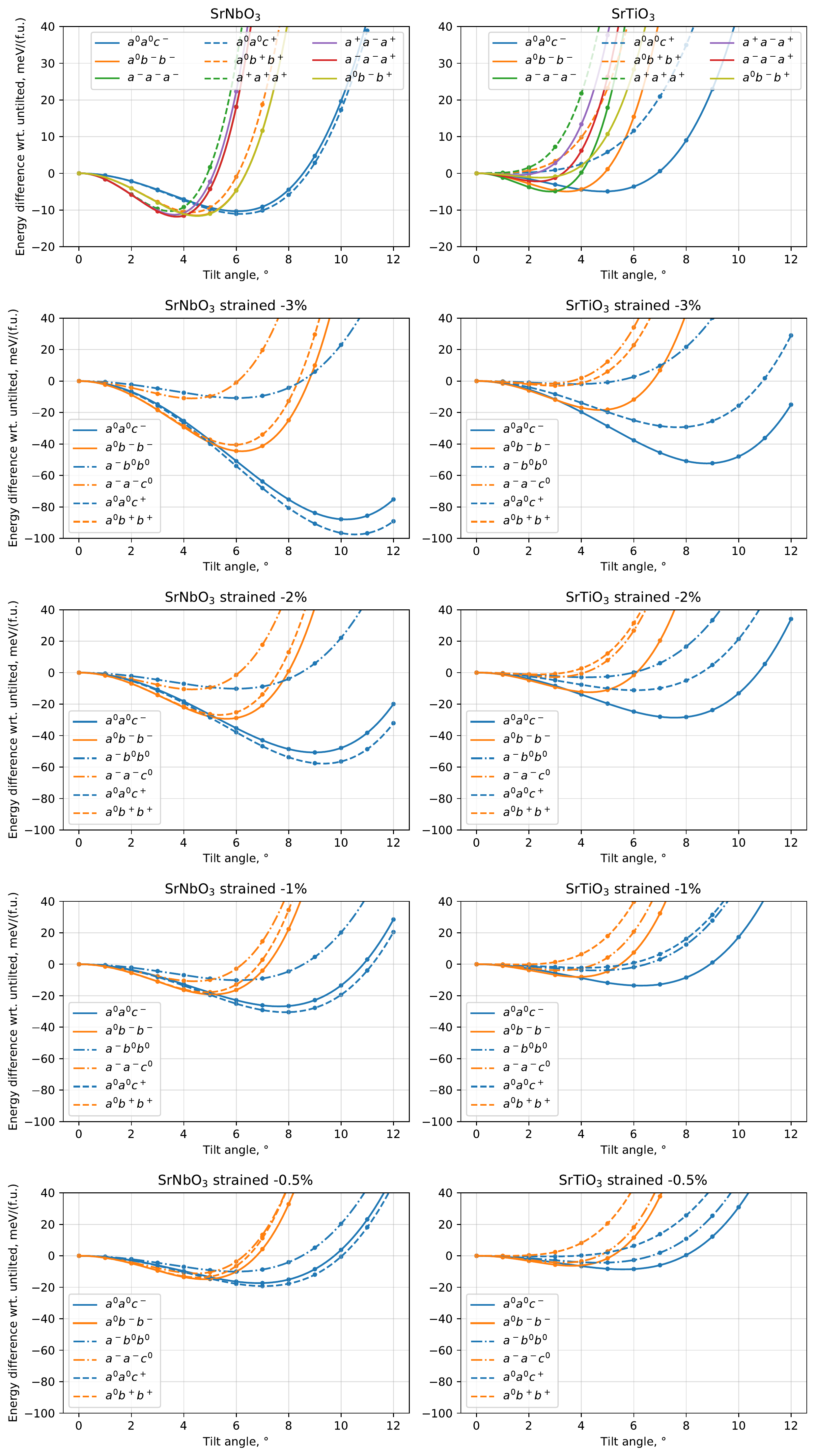}
    \caption{}
    \label{fig:tilts_appendix}
\end{figure*}

\begin{figure*}[ht!]
    \centering
    \includegraphics[trim={0 0 0 36cm},clip,height=0.85\textheight]{energies_angles.pdf}
    \caption{}
    \label{fig:tilts_appendix_next}
\end{figure*}

\clearpage
\section{All bandstructures}\label{app:allbs}
\begin{figure*}[ht!]
    \centering
    \includegraphics[width = \textwidth, angle=-90]{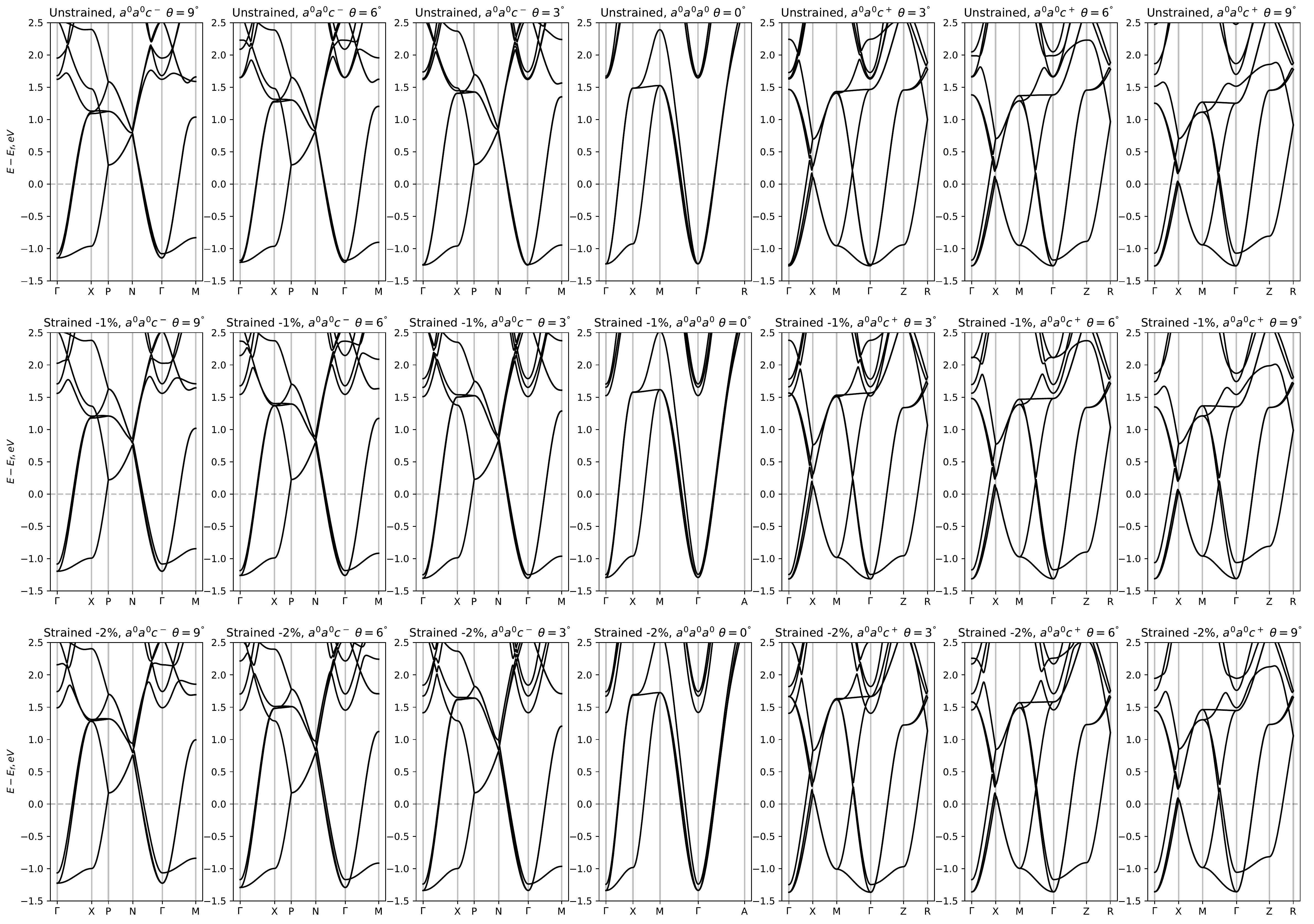}
    \caption{}
    \label{fig:bs_appendix}
\end{figure*}


\begin{thebibliography}{40}%
\makeatletter
\providecommand \@ifxundefined [1]{%
 \@ifx{#1\undefined}
}%
\providecommand \@ifnum [1]{%
 \ifnum #1\expandafter \@firstoftwo
 \else \expandafter \@secondoftwo
 \fi
}%
\providecommand \@ifx [1]{%
 \ifx #1\expandafter \@firstoftwo
 \else \expandafter \@secondoftwo
 \fi
}%
\providecommand \natexlab [1]{#1}%
\providecommand \enquote  [1]{``#1''}%
\providecommand \bibnamefont  [1]{#1}%
\providecommand \bibfnamefont [1]{#1}%
\providecommand \citenamefont [1]{#1}%
\providecommand \href@noop [0]{\@secondoftwo}%
\providecommand \href [0]{\begingroup \@sanitize@url \@href}%
\providecommand \@href[1]{\@@startlink{#1}\@@href}%
\providecommand \@@href[1]{\endgroup#1\@@endlink}%
\providecommand \@sanitize@url [0]{\catcode `\\12\catcode `\$12\catcode
  `\&12\catcode `\#12\catcode `\^12\catcode `\_12\catcode `\%12\relax}%
\providecommand \@@startlink[1]{}%
\providecommand \@@endlink[0]{}%
\providecommand \url  [0]{\begingroup\@sanitize@url \@url }%
\providecommand \@url [1]{\endgroup\@href {#1}{\urlprefix }}%
\providecommand \urlprefix  [0]{URL }%
\providecommand \Eprint [0]{\href }%
\providecommand \doibase [0]{https://doi.org/}%
\providecommand \selectlanguage [0]{\@gobble}%
\providecommand \bibinfo  [0]{\@secondoftwo}%
\providecommand \bibfield  [0]{\@secondoftwo}%
\providecommand \translation [1]{[#1]}%
\providecommand \BibitemOpen [0]{}%
\providecommand \bibitemStop [0]{}%
\providecommand \bibitemNoStop [0]{.\EOS\space}%
\providecommand \EOS [0]{\spacefactor3000\relax}%
\providecommand \BibitemShut  [1]{\csname bibitem#1\endcsname}%
\let\auto@bib@innerbib\@empty
%</preamble>
\bibitem [{\citenamefont {Bhalla}\ \emph {et~al.}(2000)\citenamefont {Bhalla},
  \citenamefont {Guo},\ and\ \citenamefont {Roy}}]{Bhalla2000}%
  \BibitemOpen
  \bibfield  {author} {\bibinfo {author} {\bibfnamefont {A.~S.}\ \bibnamefont
  {Bhalla}}, \bibinfo {author} {\bibfnamefont {R.}~\bibnamefont {Guo}},\ and\
  \bibinfo {author} {\bibfnamefont {R.}~\bibnamefont {Roy}},\ }\href
  {https://doi.org/10.1007/s100190000062} {\bibinfo {title} {{The perovskite
  structure - A review of its role in ceramic science and technology}}}
  (\bibinfo {year} {2000})\BibitemShut {NoStop}%
\bibitem [{\citenamefont {Panda}\ and\ \citenamefont
  {Sahoo}(2015)}]{Panda2015}%
  \BibitemOpen
  \bibfield  {author} {\bibinfo {author} {\bibfnamefont {P.~K.}\ \bibnamefont
  {Panda}}\ and\ \bibinfo {author} {\bibfnamefont {B.}~\bibnamefont {Sahoo}},\
  }\bibfield  {title} {\bibinfo {title} {{PZT to lead free piezo ceramics: A
  review}},\ }\bibfield  {journal} {\bibinfo  {journal} {Ferroelectrics}\
  }\href {https://doi.org/10.1080/00150193.2015.997146}
  {10.1080/00150193.2015.997146} (\bibinfo {year} {2015})\BibitemShut {NoStop}%
\bibitem [{\citenamefont {Yin}\ \emph {et~al.}(2017)\citenamefont {Yin},
  \citenamefont {Tudu},\ and\ \citenamefont {Tiwari}}]{Yin2017}%
  \BibitemOpen
  \bibfield  {author} {\bibinfo {author} {\bibfnamefont {Y.}~\bibnamefont
  {Yin}}, \bibinfo {author} {\bibfnamefont {B.}~\bibnamefont {Tudu}},\ and\
  \bibinfo {author} {\bibfnamefont {A.}~\bibnamefont {Tiwari}},\ }\bibfield
  {title} {\bibinfo {title} {{Recent advances in oxide thermoelectric materials
  and modules}},\ }\bibfield  {journal} {\bibinfo  {journal} {Vacuum}\ }\href
  {https://doi.org/10.1016/j.vacuum.2017.04.015} {10.1016/j.vacuum.2017.04.015}
  (\bibinfo {year} {2017})\BibitemShut {NoStop}%
\bibitem [{\citenamefont {Sunarso}\ \emph {et~al.}(2017)\citenamefont
  {Sunarso}, \citenamefont {Hashim}, \citenamefont {Zhu},\ and\ \citenamefont
  {Zhou}}]{Sunarso2017}%
  \BibitemOpen
  \bibfield  {author} {\bibinfo {author} {\bibfnamefont {J.}~\bibnamefont
  {Sunarso}}, \bibinfo {author} {\bibfnamefont {S.~S.}\ \bibnamefont {Hashim}},
  \bibinfo {author} {\bibfnamefont {N.}~\bibnamefont {Zhu}},\ and\ \bibinfo
  {author} {\bibfnamefont {W.}~\bibnamefont {Zhou}},\ }\href
  {https://doi.org/10.1016/j.pecs.2017.03.003} {\bibinfo {title} {{Perovskite
  oxides applications in high temperature oxygen separation, solid oxide fuel
  cell and membrane reactor: A review}}} (\bibinfo {year} {2017})\BibitemShut
  {NoStop}%
\bibitem [{\citenamefont {Schooley}\ \emph {et~al.}(1964)\citenamefont
  {Schooley}, \citenamefont {Hosler},\ and\ \citenamefont
  {Cohen}}]{Schooley1964}%
  \BibitemOpen
  \bibfield  {author} {\bibinfo {author} {\bibfnamefont {J.~F.}\ \bibnamefont
  {Schooley}}, \bibinfo {author} {\bibfnamefont {W.~R.}\ \bibnamefont
  {Hosler}},\ and\ \bibinfo {author} {\bibfnamefont {M.~L.}\ \bibnamefont
  {Cohen}},\ }\bibfield  {title} {\bibinfo {title} {{Superconductivity in
  semiconducting SrTiO3}},\ }\bibfield  {journal} {\bibinfo  {journal}
  {Physical Review Letters}\ }\href
  {https://doi.org/10.1103/PhysRevLett.12.474} {10.1103/PhysRevLett.12.474}
  (\bibinfo {year} {1964})\BibitemShut {NoStop}%
\bibitem [{\citenamefont {Baldini}\ \emph {et~al.}(2015)\citenamefont
  {Baldini}, \citenamefont {Muramatsu}, \citenamefont {Sherafati},
  \citenamefont {kwang Mao}, \citenamefont {Malavasi}, \citenamefont
  {Postorino}, \citenamefont {Satpathy},\ and\ \citenamefont
  {Struzhkin}}]{Baldini_CMR}%
  \BibitemOpen
  \bibfield  {author} {\bibinfo {author} {\bibfnamefont {M.}~\bibnamefont
  {Baldini}}, \bibinfo {author} {\bibfnamefont {T.}~\bibnamefont {Muramatsu}},
  \bibinfo {author} {\bibfnamefont {M.}~\bibnamefont {Sherafati}}, \bibinfo
  {author} {\bibfnamefont {H.}~\bibnamefont {kwang Mao}}, \bibinfo {author}
  {\bibfnamefont {L.}~\bibnamefont {Malavasi}}, \bibinfo {author}
  {\bibfnamefont {P.}~\bibnamefont {Postorino}}, \bibinfo {author}
  {\bibfnamefont {S.}~\bibnamefont {Satpathy}},\ and\ \bibinfo {author}
  {\bibfnamefont {V.~V.}\ \bibnamefont {Struzhkin}},\ }\bibfield  {title}
  {\bibinfo {title} {{Origin of colossal magnetoresistance in LaMnO3
  manganite}},\ }\href {https://doi.org/10.1073/pnas.1424866112} {\bibfield
  {journal} {\bibinfo  {journal} {Proceedings of the National Academy of
  Sciences}\ }\textbf {\bibinfo {volume} {112}},\ \bibinfo {pages} {10869}
  (\bibinfo {year} {2015})},\ \Eprint
  {https://arxiv.org/abs/https://www.pnas.org/doi/pdf/10.1073/pnas.1424866112}
  {https://www.pnas.org/doi/pdf/10.1073/pnas.1424866112} \BibitemShut {NoStop}%
\bibitem [{\citenamefont {Wong}\ \emph {et~al.}(2010)\citenamefont {Wong},
  \citenamefont {Baek}, \citenamefont {Chopdekar}, \citenamefont {Mehta},
  \citenamefont {Jang}, \citenamefont {Eom},\ and\ \citenamefont
  {Suzuki}}]{Wong2010}%
  \BibitemOpen
  \bibfield  {author} {\bibinfo {author} {\bibfnamefont {F.~J.}\ \bibnamefont
  {Wong}}, \bibinfo {author} {\bibfnamefont {S.~H.}\ \bibnamefont {Baek}},
  \bibinfo {author} {\bibfnamefont {R.~V.}\ \bibnamefont {Chopdekar}}, \bibinfo
  {author} {\bibfnamefont {V.~V.}\ \bibnamefont {Mehta}}, \bibinfo {author}
  {\bibfnamefont {H.~W.}\ \bibnamefont {Jang}}, \bibinfo {author}
  {\bibfnamefont {C.~B.}\ \bibnamefont {Eom}},\ and\ \bibinfo {author}
  {\bibfnamefont {Y.}~\bibnamefont {Suzuki}},\ }\bibfield  {title} {\bibinfo
  {title} {{Metallicity in LaTiO3 thin films induced by lattice deformation}},\
  }\bibfield  {journal} {\bibinfo  {journal} {Physical Review B - Condensed
  Matter and Materials Physics}\ }\href
  {https://doi.org/10.1103/PhysRevB.81.161101} {10.1103/PhysRevB.81.161101}
  (\bibinfo {year} {2010})\BibitemShut {NoStop}%
\bibitem [{\citenamefont {Gu}\ \emph {et~al.}(2013)\citenamefont {Gu},
  \citenamefont {Laverock}, \citenamefont {Chen}, \citenamefont {Smith},
  \citenamefont {Wolf},\ and\ \citenamefont {Lu}}]{Gu2013}%
  \BibitemOpen
  \bibfield  {author} {\bibinfo {author} {\bibfnamefont {M.}~\bibnamefont
  {Gu}}, \bibinfo {author} {\bibfnamefont {J.}~\bibnamefont {Laverock}},
  \bibinfo {author} {\bibfnamefont {B.}~\bibnamefont {Chen}}, \bibinfo {author}
  {\bibfnamefont {K.~E.}\ \bibnamefont {Smith}}, \bibinfo {author}
  {\bibfnamefont {S.~A.}\ \bibnamefont {Wolf}},\ and\ \bibinfo {author}
  {\bibfnamefont {J.}~\bibnamefont {Lu}},\ }\bibfield  {title} {\bibinfo
  {title} {{Metal-insulator transition induced in CaVO3 thin films}},\ }in\
  \href {https://doi.org/10.1063/1.4798963} {\emph {\bibinfo {booktitle}
  {Journal of Applied Physics}}}\ (\bibinfo {year} {2013})\BibitemShut
  {NoStop}%
\bibitem [{\citenamefont {Ok}\ \emph {et~al.}(2021)\citenamefont {Ok},
  \citenamefont {Mohanta}, \citenamefont {Zhang}, \citenamefont {Yoon},
  \citenamefont {Okamoto}, \citenamefont {Choi}, \citenamefont {Zhou},
  \citenamefont {Briggeman}, \citenamefont {Irvin}, \citenamefont {Lupini},
  \citenamefont {Pai}, \citenamefont {Skoropata}, \citenamefont {Sohn},
  \citenamefont {Li}, \citenamefont {Miao}, \citenamefont {Lawrie},
  \citenamefont {Choi}, \citenamefont {Eres}, \citenamefont {Levy},\ and\
  \citenamefont {Lee}}]{Ok2021}%
  \BibitemOpen
  \bibfield  {author} {\bibinfo {author} {\bibfnamefont {J.~M.}\ \bibnamefont
  {Ok}}, \bibinfo {author} {\bibfnamefont {N.}~\bibnamefont {Mohanta}},
  \bibinfo {author} {\bibfnamefont {J.}~\bibnamefont {Zhang}}, \bibinfo
  {author} {\bibfnamefont {S.}~\bibnamefont {Yoon}}, \bibinfo {author}
  {\bibfnamefont {S.}~\bibnamefont {Okamoto}}, \bibinfo {author} {\bibfnamefont
  {E.~S.}\ \bibnamefont {Choi}}, \bibinfo {author} {\bibfnamefont
  {H.}~\bibnamefont {Zhou}}, \bibinfo {author} {\bibfnamefont {M.}~\bibnamefont
  {Briggeman}}, \bibinfo {author} {\bibfnamefont {P.}~\bibnamefont {Irvin}},
  \bibinfo {author} {\bibfnamefont {A.~R.}\ \bibnamefont {Lupini}}, \bibinfo
  {author} {\bibfnamefont {Y.~Y.}\ \bibnamefont {Pai}}, \bibinfo {author}
  {\bibfnamefont {E.}~\bibnamefont {Skoropata}}, \bibinfo {author}
  {\bibfnamefont {C.}~\bibnamefont {Sohn}}, \bibinfo {author} {\bibfnamefont
  {H.}~\bibnamefont {Li}}, \bibinfo {author} {\bibfnamefont {H.}~\bibnamefont
  {Miao}}, \bibinfo {author} {\bibfnamefont {B.}~\bibnamefont {Lawrie}},
  \bibinfo {author} {\bibfnamefont {W.~S.}\ \bibnamefont {Choi}}, \bibinfo
  {author} {\bibfnamefont {G.}~\bibnamefont {Eres}}, \bibinfo {author}
  {\bibfnamefont {J.}~\bibnamefont {Levy}},\ and\ \bibinfo {author}
  {\bibfnamefont {H.~N.}\ \bibnamefont {Lee}},\ }\bibfield  {title} {\bibinfo
  {title} {{Correlated oxide Dirac semimetal in the extreme quantum limit}},\
  }\bibfield  {journal} {\bibinfo  {journal} {Science Advances}\ }\textbf
  {\bibinfo {volume} {7}},\ \href {https://doi.org/10.1126/sciadv.abf9631}
  {10.1126/sciadv.abf9631} (\bibinfo {year} {2021})\BibitemShut {NoStop}%
\bibitem [{\citenamefont {Rondinelli}\ and\ \citenamefont
  {Spaldin}(2011)}]{Rondinelli2011}%
  \BibitemOpen
  \bibfield  {author} {\bibinfo {author} {\bibfnamefont {J.~M.}\ \bibnamefont
  {Rondinelli}}\ and\ \bibinfo {author} {\bibfnamefont {N.~A.}\ \bibnamefont
  {Spaldin}},\ }\href {https://doi.org/10.1002/adma.201101152} {\bibinfo
  {title} {{Structure and properties of functional oxide thin films: Insights
  from electronic-structure calculations}}} (\bibinfo {year}
  {2011})\BibitemShut {NoStop}%
\bibitem [{\citenamefont {Rondinelli}\ \emph {et~al.}(2012)\citenamefont
  {Rondinelli}, \citenamefont {May},\ and\ \citenamefont
  {Freeland}}]{Rondinelli2012}%
  \BibitemOpen
  \bibfield  {author} {\bibinfo {author} {\bibfnamefont {J.~M.}\ \bibnamefont
  {Rondinelli}}, \bibinfo {author} {\bibfnamefont {S.~J.}\ \bibnamefont
  {May}},\ and\ \bibinfo {author} {\bibfnamefont {J.~W.}\ \bibnamefont
  {Freeland}},\ }\bibfield  {title} {\bibinfo {title} {{Control of octahedral
  connectivity in perovskite oxide heterostructures: An emerging route to
  multifunctional materials discovery}},\ }\bibfield  {journal} {\bibinfo
  {journal} {MRS Bulletin}\ }\textbf {\bibinfo {volume} {37}},\ \href
  {https://doi.org/10.1557/mrs.2012.49} {10.1557/mrs.2012.49} (\bibinfo {year}
  {2012})\BibitemShut {NoStop}%
\bibitem [{\citenamefont {Oka}\ \emph {et~al.}(2015)\citenamefont {Oka},
  \citenamefont {Hirose}, \citenamefont {Nakao}, \citenamefont {Fukumura},\
  and\ \citenamefont {Hasegawa}}]{Oka2015}%
  \BibitemOpen
  \bibfield  {author} {\bibinfo {author} {\bibfnamefont {D.}~\bibnamefont
  {Oka}}, \bibinfo {author} {\bibfnamefont {Y.}~\bibnamefont {Hirose}},
  \bibinfo {author} {\bibfnamefont {S.}~\bibnamefont {Nakao}}, \bibinfo
  {author} {\bibfnamefont {T.}~\bibnamefont {Fukumura}},\ and\ \bibinfo
  {author} {\bibfnamefont {T.}~\bibnamefont {Hasegawa}},\ }\bibfield  {title}
  {\bibinfo {title} {{Intrinsic high electrical conductivity of stoichiometric
  SrNbO3 epitaxial thin films}},\ }\bibfield  {journal} {\bibinfo  {journal}
  {Physical Review B - Condensed Matter and Materials Physics}\ }\href
  {https://doi.org/10.1103/PhysRevB.92.205102} {10.1103/PhysRevB.92.205102}
  (\bibinfo {year} {2015})\BibitemShut {NoStop}%
\bibitem [{\citenamefont {Park}\ \emph {et~al.}(2020)\citenamefont {Park},
  \citenamefont {Roth}, \citenamefont {Oka}, \citenamefont {Hirose},
  \citenamefont {Hasegawa}, \citenamefont {Paul}, \citenamefont {Pogrebnyakov},
  \citenamefont {Gopalan}, \citenamefont {Birol},\ and\ \citenamefont
  {Engel-Herbert}}]{Park2020}%
  \BibitemOpen
  \bibfield  {author} {\bibinfo {author} {\bibfnamefont {Y.}~\bibnamefont
  {Park}}, \bibinfo {author} {\bibfnamefont {J.}~\bibnamefont {Roth}}, \bibinfo
  {author} {\bibfnamefont {D.}~\bibnamefont {Oka}}, \bibinfo {author}
  {\bibfnamefont {Y.}~\bibnamefont {Hirose}}, \bibinfo {author} {\bibfnamefont
  {T.}~\bibnamefont {Hasegawa}}, \bibinfo {author} {\bibfnamefont
  {A.}~\bibnamefont {Paul}}, \bibinfo {author} {\bibfnamefont {A.}~\bibnamefont
  {Pogrebnyakov}}, \bibinfo {author} {\bibfnamefont {V.}~\bibnamefont
  {Gopalan}}, \bibinfo {author} {\bibfnamefont {T.}~\bibnamefont {Birol}},\
  and\ \bibinfo {author} {\bibfnamefont {R.}~\bibnamefont {Engel-Herbert}},\
  }\bibfield  {title} {\bibinfo {title} {{SrNbO3 as a transparent conductor in
  the visible and ultraviolet spectra}},\ }\bibfield  {journal} {\bibinfo
  {journal} {Communications Physics}\ }\textbf {\bibinfo {volume} {3}},\ \href
  {https://doi.org/10.1038/s42005-020-0372-9} {10.1038/s42005-020-0372-9}
  (\bibinfo {year} {2020})\BibitemShut {NoStop}%
\bibitem [{\citenamefont {Zhao}\ \emph {et~al.}(2021)\citenamefont {Zhao},
  \citenamefont {Wang}, \citenamefont {Malyi},\ and\ \citenamefont
  {Zunger}}]{Zhao2021}%
  \BibitemOpen
  \bibfield  {author} {\bibinfo {author} {\bibfnamefont {X.~G.}\ \bibnamefont
  {Zhao}}, \bibinfo {author} {\bibfnamefont {Z.}~\bibnamefont {Wang}}, \bibinfo
  {author} {\bibfnamefont {O.~I.}\ \bibnamefont {Malyi}},\ and\ \bibinfo
  {author} {\bibfnamefont {A.}~\bibnamefont {Zunger}},\ }\bibfield  {title}
  {\bibinfo {title} {{Effect of static local distortions vs. dynamic motions on
  the stability and band gaps of cubic oxide and halide perovskites}},\
  }\bibfield  {journal} {\bibinfo  {journal} {Materials Today}\ }\textbf
  {\bibinfo {volume} {49}},\ \href
  {https://doi.org/10.1016/j.mattod.2021.05.021} {10.1016/j.mattod.2021.05.021}
  (\bibinfo {year} {2021})\BibitemShut {NoStop}%
\bibitem [{\citenamefont {Varignon}\ \emph {et~al.}(2019)\citenamefont
  {Varignon}, \citenamefont {Bibes},\ and\ \citenamefont
  {Zunger}}]{Varignon2019}%
  \BibitemOpen
  \bibfield  {author} {\bibinfo {author} {\bibfnamefont {J.}~\bibnamefont
  {Varignon}}, \bibinfo {author} {\bibfnamefont {M.}~\bibnamefont {Bibes}},\
  and\ \bibinfo {author} {\bibfnamefont {A.}~\bibnamefont {Zunger}},\
  }\bibfield  {title} {\bibinfo {title} {{Origin of band gaps in 3d perovskite
  oxides}},\ }\bibfield  {journal} {\bibinfo  {journal} {Nature
  Communications}\ }\textbf {\bibinfo {volume} {10}},\ \href
  {https://doi.org/10.1038/s41467-019-09698-6} {10.1038/s41467-019-09698-6}
  (\bibinfo {year} {2019})\BibitemShut {NoStop}%
\bibitem [{\citenamefont {Mott}(2004)}]{Mott2004}%
  \BibitemOpen
  \bibfield  {author} {\bibinfo {author} {\bibfnamefont {N.}~\bibnamefont
  {Mott}},\ }\href {https://doi.org/10.1201/b12795} {\emph {\bibinfo {title}
  {{Metal-Insulator Transitions}}}}\ (\bibinfo  {publisher} {CRC Press},\
  \bibinfo {year} {2004})\BibitemShut {NoStop}%
\bibitem [{\citenamefont {Mirjolet}\ \emph {et~al.}(2021)\citenamefont
  {Mirjolet}, \citenamefont {Kataja}, \citenamefont {Hakala}, \citenamefont
  {Komissinskiy}, \citenamefont {Alff}, \citenamefont {Herranz},\ and\
  \citenamefont {Fontcuberta}}]{Mirjolet2021}%
  \BibitemOpen
  \bibfield  {author} {\bibinfo {author} {\bibfnamefont {M.}~\bibnamefont
  {Mirjolet}}, \bibinfo {author} {\bibfnamefont {M.}~\bibnamefont {Kataja}},
  \bibinfo {author} {\bibfnamefont {T.~K.}\ \bibnamefont {Hakala}}, \bibinfo
  {author} {\bibfnamefont {P.}~\bibnamefont {Komissinskiy}}, \bibinfo {author}
  {\bibfnamefont {L.}~\bibnamefont {Alff}}, \bibinfo {author} {\bibfnamefont
  {G.}~\bibnamefont {Herranz}},\ and\ \bibinfo {author} {\bibfnamefont
  {J.}~\bibnamefont {Fontcuberta}},\ }\bibfield  {title} {\bibinfo {title}
  {{Optical Plasmon Excitation in Transparent Conducting SrNbO3 and SrVO3 Thin
  Films}},\ }\bibfield  {journal} {\bibinfo  {journal} {Advanced Optical
  Materials}\ }\textbf {\bibinfo {volume} {9}},\ \href
  {https://doi.org/10.1002/adom.202100520} {10.1002/adom.202100520} (\bibinfo
  {year} {2021})\BibitemShut {NoStop}%
\bibitem [{\citenamefont {Kohn}\ and\ \citenamefont {Sham}(1965)}]{Kohn1965}%
  \BibitemOpen
  \bibfield  {author} {\bibinfo {author} {\bibfnamefont {W.}~\bibnamefont
  {Kohn}}\ and\ \bibinfo {author} {\bibfnamefont {L.~J.}\ \bibnamefont
  {Sham}},\ }\bibfield  {title} {\bibinfo {title} {{Self-consistent equations
  including exchange and correlation effects}},\ }\bibfield  {journal}
  {\bibinfo  {journal} {Physical Review}\ }\textbf {\bibinfo {volume} {140}},\
  \href {https://doi.org/10.1103/PhysRev.140.A1133} {10.1103/PhysRev.140.A1133}
  (\bibinfo {year} {1965})\BibitemShut {NoStop}%
\bibitem [{\citenamefont {Bl{\"{o}}chl}(1994)}]{Blochl1994}%
  \BibitemOpen
  \bibfield  {author} {\bibinfo {author} {\bibfnamefont {P.~E.}\ \bibnamefont
  {Bl{\"{o}}chl}},\ }\bibfield  {title} {\bibinfo {title} {{Projector
  augmented-wave method}},\ }\bibfield  {journal} {\bibinfo  {journal}
  {Physical Review B}\ }\textbf {\bibinfo {volume} {50}},\ \href
  {https://doi.org/10.1103/PhysRevB.50.17953} {10.1103/PhysRevB.50.17953}
  (\bibinfo {year} {1994})\BibitemShut {NoStop}%
\bibitem [{\citenamefont {Joubert}(1999)}]{Joubert1999}%
  \BibitemOpen
  \bibfield  {author} {\bibinfo {author} {\bibfnamefont {D.}~\bibnamefont
  {Joubert}},\ }\bibfield  {title} {\bibinfo {title} {{From ultrasoft
  pseudopotentials to the projector augmented-wave method}},\ }\bibfield
  {journal} {\bibinfo  {journal} {Physical Review B - Condensed Matter and
  Materials Physics}\ }\textbf {\bibinfo {volume} {59}},\ \href
  {https://doi.org/10.1103/PhysRevB.59.1758} {10.1103/PhysRevB.59.1758}
  (\bibinfo {year} {1999})\BibitemShut {NoStop}%
\bibitem [{\citenamefont {Kresse}\ and\ \citenamefont
  {Furthm{\"{u}}ller}(1996)}]{Kresse1996}%
  \BibitemOpen
  \bibfield  {author} {\bibinfo {author} {\bibfnamefont {G.}~\bibnamefont
  {Kresse}}\ and\ \bibinfo {author} {\bibfnamefont {J.}~\bibnamefont
  {Furthm{\"{u}}ller}},\ }\bibfield  {title} {\bibinfo {title} {{Efficient
  iterative schemes for ab initio total-energy calculations using a plane-wave
  basis set}},\ }\bibfield  {journal} {\bibinfo  {journal} {Physical Review B -
  Condensed Matter and Materials Physics}\ }\textbf {\bibinfo {volume} {54}},\
  \href {https://doi.org/10.1103/PhysRevB.54.11169} {10.1103/PhysRevB.54.11169}
  (\bibinfo {year} {1996})\BibitemShut {NoStop}%
\bibitem [{\citenamefont {Perdew}\ \emph {et~al.}(2008)\citenamefont {Perdew},
  \citenamefont {Ruzsinszky}, \citenamefont {Csonka}, \citenamefont {Vydrov},
  \citenamefont {Scuseria}, \citenamefont {Constantin}, \citenamefont {Zhou},\
  and\ \citenamefont {Burke}}]{Perdew2008}%
  \BibitemOpen
  \bibfield  {author} {\bibinfo {author} {\bibfnamefont {J.~P.}\ \bibnamefont
  {Perdew}}, \bibinfo {author} {\bibfnamefont {A.}~\bibnamefont {Ruzsinszky}},
  \bibinfo {author} {\bibfnamefont {G.~I.}\ \bibnamefont {Csonka}}, \bibinfo
  {author} {\bibfnamefont {O.~A.}\ \bibnamefont {Vydrov}}, \bibinfo {author}
  {\bibfnamefont {G.~E.}\ \bibnamefont {Scuseria}}, \bibinfo {author}
  {\bibfnamefont {L.~A.}\ \bibnamefont {Constantin}}, \bibinfo {author}
  {\bibfnamefont {X.}~\bibnamefont {Zhou}},\ and\ \bibinfo {author}
  {\bibfnamefont {K.}~\bibnamefont {Burke}},\ }\bibfield  {title} {\bibinfo
  {title} {{Restoring the density-gradient expansion for exchange in solids and
  surfaces}},\ }\bibfield  {journal} {\bibinfo  {journal} {Physical Review
  Letters}\ }\href {https://doi.org/10.1103/PhysRevLett.100.136406}
  {10.1103/PhysRevLett.100.136406} (\bibinfo {year} {2008}),\ \Eprint
  {https://arxiv.org/abs/0711.0156} {arXiv:0711.0156} \BibitemShut {NoStop}%
\bibitem [{\citenamefont {Aschauer}\ and\ \citenamefont
  {Spaldin}(2014)}]{Aschauer2014}%
  \BibitemOpen
  \bibfield  {author} {\bibinfo {author} {\bibfnamefont {U.}~\bibnamefont
  {Aschauer}}\ and\ \bibinfo {author} {\bibfnamefont {N.~A.}\ \bibnamefont
  {Spaldin}},\ }\bibfield  {title} {\bibinfo {title} {{Competition and
  cooperation between antiferrodistortive and ferroelectric instabilities in
  the model perovskite SrTiO3}},\ }\bibfield  {journal} {\bibinfo  {journal}
  {Journal of Physics Condensed Matter}\ }\textbf {\bibinfo {volume} {26}},\
  \href {https://doi.org/10.1088/0953-8984/26/12/122203}
  {10.1088/0953-8984/26/12/122203} (\bibinfo {year} {2014})\BibitemShut
  {NoStop}%
\bibitem [{\citenamefont {Haule}\ \emph {et~al.}(2010)\citenamefont {Haule},
  \citenamefont {Yee},\ and\ \citenamefont {Kim}}]{DMFTimple}%
  \BibitemOpen
  \bibfield  {author} {\bibinfo {author} {\bibfnamefont {K.}~\bibnamefont
  {Haule}}, \bibinfo {author} {\bibfnamefont {C.-H.}\ \bibnamefont {Yee}},\
  and\ \bibinfo {author} {\bibfnamefont {K.}~\bibnamefont {Kim}},\ }\bibfield
  {title} {\bibinfo {title} {Dynamical mean-field theory within the
  full-potential methods: Electronic structure of ${\text{ceirin}}_{5}$,
  ${\text{cecoin}}_{5}$, and ${\text{cerhin}}_{5}$},\ }\href
  {https://doi.org/10.1103/PhysRevB.81.195107} {\bibfield  {journal} {\bibinfo
  {journal} {Phys. Rev. B}\ }\textbf {\bibinfo {volume} {81}},\ \bibinfo
  {pages} {195107} (\bibinfo {year} {2010})}\BibitemShut {NoStop}%
\bibitem [{\citenamefont {Perdew}\ \emph {et~al.}(1996)\citenamefont {Perdew},
  \citenamefont {Burke},\ and\ \citenamefont {Ernzerhof}}]{pbe}%
  \BibitemOpen
  \bibfield  {author} {\bibinfo {author} {\bibfnamefont {J.~P.}\ \bibnamefont
  {Perdew}}, \bibinfo {author} {\bibfnamefont {K.}~\bibnamefont {Burke}},\ and\
  \bibinfo {author} {\bibfnamefont {M.}~\bibnamefont {Ernzerhof}},\ }\bibfield
  {title} {\bibinfo {title} {Generalized gradient approximation made simple},\
  }\href {https://doi.org/10.1103/PhysRevLett.77.3865} {\bibfield  {journal}
  {\bibinfo  {journal} {Phys. Rev. Lett.}\ }\textbf {\bibinfo {volume} {77}},\
  \bibinfo {pages} {3865} (\bibinfo {year} {1996})}\BibitemShut {NoStop}%
\bibitem [{\citenamefont {Blaha}\ \emph {et~al.}(2001)\citenamefont {Blaha},
  \citenamefont {Madsen}, \citenamefont {Kvasnicka},\ and\ \citenamefont
  {Luitz}}]{wien}%
  \BibitemOpen
  \bibfield  {author} {\bibinfo {author} {\bibfnamefont {P.}~\bibnamefont
  {Blaha}}, \bibinfo {author} {\bibfnamefont {G.~K.~H.}\ \bibnamefont
  {Madsen}}, \bibinfo {author} {\bibfnamefont {D.}~\bibnamefont {Kvasnicka}},\
  and\ \bibinfo {author} {\bibfnamefont {J.}~\bibnamefont {Luitz}},\
  }\href@noop {} {\emph {\bibinfo {title} {\emph{WIEN2K, An Augmented Plane
  Wave + Local Orbitals Program for Calculating Crystal Properties}}}}\
  (\bibinfo  {publisher} {(Karlheinz Schwarz, Techn. Universität Wien,
  Austria)},\ \bibinfo {year} {2001})\BibitemShut {NoStop}%
\bibitem [{\citenamefont {Haule}(2007)}]{ctqmc}%
  \BibitemOpen
  \bibfield  {author} {\bibinfo {author} {\bibfnamefont {K.}~\bibnamefont
  {Haule}},\ }\bibfield  {title} {\bibinfo {title} {Quantum monte carlo
  impurity solver for cluster dynamical mean-field theory and electronic
  structure calculations with adjustable cluster base},\ }\href
  {https://doi.org/10.1103/PhysRevB.75.155113} {\bibfield  {journal} {\bibinfo
  {journal} {Phys. Rev. B}\ }\textbf {\bibinfo {volume} {75}},\ \bibinfo
  {pages} {155113} (\bibinfo {year} {2007})}\BibitemShut {NoStop}%
\bibitem [{\citenamefont {Madsen}\ \emph {et~al.}(2018)\citenamefont {Madsen},
  \citenamefont {Carrete},\ and\ \citenamefont {Verstraete}}]{Madsen2018}%
  \BibitemOpen
  \bibfield  {author} {\bibinfo {author} {\bibfnamefont {G.~K.}\ \bibnamefont
  {Madsen}}, \bibinfo {author} {\bibfnamefont {J.}~\bibnamefont {Carrete}},\
  and\ \bibinfo {author} {\bibfnamefont {M.~J.}\ \bibnamefont {Verstraete}},\
  }\bibfield  {title} {\bibinfo {title} {{BoltzTraP2, a program for
  interpolating band structures and calculating semi-classical transport
  coefficients}},\ }\bibfield  {journal} {\bibinfo  {journal} {Computer Physics
  Communications}\ }\textbf {\bibinfo {volume} {231}},\ \href
  {https://doi.org/10.1016/j.cpc.2018.05.010} {10.1016/j.cpc.2018.05.010}
  (\bibinfo {year} {2018})\BibitemShut {NoStop}%
\bibitem [{\citenamefont {Gajdo{\v{s}}}\ \emph {et~al.}(2006)\citenamefont
  {Gajdo{\v{s}}}, \citenamefont {Hummer}, \citenamefont {Kresse}, \citenamefont
  {Furthm{\"{u}}ller},\ and\ \citenamefont {Bechstedt}}]{Gajdos2006}%
  \BibitemOpen
  \bibfield  {author} {\bibinfo {author} {\bibfnamefont {M.}~\bibnamefont
  {Gajdo{\v{s}}}}, \bibinfo {author} {\bibfnamefont {K.}~\bibnamefont
  {Hummer}}, \bibinfo {author} {\bibfnamefont {G.}~\bibnamefont {Kresse}},
  \bibinfo {author} {\bibfnamefont {J.}~\bibnamefont {Furthm{\"{u}}ller}},\
  and\ \bibinfo {author} {\bibfnamefont {F.}~\bibnamefont {Bechstedt}},\
  }\bibfield  {title} {\bibinfo {title} {{Linear optical properties in the
  projector-augmented wave methodology}},\ }\bibfield  {journal} {\bibinfo
  {journal} {Physical Review B - Condensed Matter and Materials Physics}\
  }\textbf {\bibinfo {volume} {73}},\ \href
  {https://doi.org/10.1103/PhysRevB.73.045112} {10.1103/PhysRevB.73.045112}
  (\bibinfo {year} {2006})\BibitemShut {NoStop}%
\bibitem [{\citenamefont {Glazer}(1972)}]{Glazer1972}%
  \BibitemOpen
  \bibfield  {author} {\bibinfo {author} {\bibfnamefont {A.~M.}\ \bibnamefont
  {Glazer}},\ }\bibfield  {title} {\bibinfo {title} {{The classification of
  tilted octahedra in perovskites}},\ }\bibfield  {journal} {\bibinfo
  {journal} {Acta Crystallographica Section B Structural Crystallography and
  Crystal Chemistry}\ }\textbf {\bibinfo {volume} {28}},\ \href
  {https://doi.org/10.1107/s0567740872007976} {10.1107/s0567740872007976}
  (\bibinfo {year} {1972})\BibitemShut {NoStop}%
\bibitem [{\citenamefont {Young}\ and\ \citenamefont
  {Rondinelli}(2016)}]{Young2016}%
  \BibitemOpen
  \bibfield  {author} {\bibinfo {author} {\bibfnamefont {J.}~\bibnamefont
  {Young}}\ and\ \bibinfo {author} {\bibfnamefont {J.~M.}\ \bibnamefont
  {Rondinelli}},\ }\bibfield  {title} {\bibinfo {title} {{Octahedral Rotation
  Preferences in Perovskite Iodides and Bromides}},\ }\bibfield  {journal}
  {\bibinfo  {journal} {Journal of Physical Chemistry Letters}\ }\textbf
  {\bibinfo {volume} {7}},\ \href {https://doi.org/10.1021/acs.jpclett.6b00094}
  {10.1021/acs.jpclett.6b00094} (\bibinfo {year} {2016})\BibitemShut {NoStop}%
\bibitem [{\citenamefont {Johnson-Wilke}\ \emph {et~al.}(2013)\citenamefont
  {Johnson-Wilke}, \citenamefont {Marincel}, \citenamefont {Zhu}, \citenamefont
  {Warusawithana}, \citenamefont {Hatt}, \citenamefont {Sayre}, \citenamefont
  {Delaney}, \citenamefont {Engel-Herbert}, \citenamefont {Schlep{\"{u}}tz},
  \citenamefont {Kim}, \citenamefont {Gopalan}, \citenamefont {Spaldin},
  \citenamefont {Schlom}, \citenamefont {Ryan},\ and\ \citenamefont
  {Trolier-Mckinstry}}]{Johnson-Wilke2013}%
  \BibitemOpen
  \bibfield  {author} {\bibinfo {author} {\bibfnamefont {R.~L.}\ \bibnamefont
  {Johnson-Wilke}}, \bibinfo {author} {\bibfnamefont {D.}~\bibnamefont
  {Marincel}}, \bibinfo {author} {\bibfnamefont {S.}~\bibnamefont {Zhu}},
  \bibinfo {author} {\bibfnamefont {M.~P.}\ \bibnamefont {Warusawithana}},
  \bibinfo {author} {\bibfnamefont {A.}~\bibnamefont {Hatt}}, \bibinfo {author}
  {\bibfnamefont {J.}~\bibnamefont {Sayre}}, \bibinfo {author} {\bibfnamefont
  {K.~T.}\ \bibnamefont {Delaney}}, \bibinfo {author} {\bibfnamefont
  {R.}~\bibnamefont {Engel-Herbert}}, \bibinfo {author} {\bibfnamefont {C.~M.}\
  \bibnamefont {Schlep{\"{u}}tz}}, \bibinfo {author} {\bibfnamefont {J.~W.}\
  \bibnamefont {Kim}}, \bibinfo {author} {\bibfnamefont {V.}~\bibnamefont
  {Gopalan}}, \bibinfo {author} {\bibfnamefont {N.~A.}\ \bibnamefont
  {Spaldin}}, \bibinfo {author} {\bibfnamefont {D.~G.}\ \bibnamefont {Schlom}},
  \bibinfo {author} {\bibfnamefont {P.~J.}\ \bibnamefont {Ryan}},\ and\
  \bibinfo {author} {\bibfnamefont {S.}~\bibnamefont {Trolier-Mckinstry}},\
  }\bibfield  {title} {\bibinfo {title} {{Quantification of octahedral
  rotations in strained LaAlO3 films via synchrotron x-ray diffraction}},\
  }\bibfield  {journal} {\bibinfo  {journal} {Physical Review B - Condensed
  Matter and Materials Physics}\ }\href
  {https://doi.org/10.1103/PhysRevB.88.174101} {10.1103/PhysRevB.88.174101}
  (\bibinfo {year} {2013})\BibitemShut {NoStop}%
\bibitem [{\citenamefont {Moreau}\ \emph {et~al.}(2017)\citenamefont {Moreau},
  \citenamefont {Marthinsen}, \citenamefont {Selbach},\ and\ \citenamefont
  {Tybell}}]{Moreau2017}%
  \BibitemOpen
  \bibfield  {author} {\bibinfo {author} {\bibfnamefont {M.}~\bibnamefont
  {Moreau}}, \bibinfo {author} {\bibfnamefont {A.}~\bibnamefont {Marthinsen}},
  \bibinfo {author} {\bibfnamefont {S.~M.}\ \bibnamefont {Selbach}},\ and\
  \bibinfo {author} {\bibfnamefont {T.}~\bibnamefont {Tybell}},\ }\bibfield
  {title} {\bibinfo {title} {{Strain-phonon coupling in (111)-oriented
  perovskite oxides}},\ }\bibfield  {journal} {\bibinfo  {journal} {Physical
  Review B}\ }\textbf {\bibinfo {volume} {96}},\ \href
  {https://doi.org/10.1103/PhysRevB.96.094109} {10.1103/PhysRevB.96.094109}
  (\bibinfo {year} {2017})\BibitemShut {NoStop}%
\bibitem [{\citenamefont {Uchida}\ \emph {et~al.}(2003)\citenamefont {Uchida},
  \citenamefont {Tsuneyuki},\ and\ \citenamefont {Schimizu}}]{Uchida2003}%
  \BibitemOpen
  \bibfield  {author} {\bibinfo {author} {\bibfnamefont {K.}~\bibnamefont
  {Uchida}}, \bibinfo {author} {\bibfnamefont {S.}~\bibnamefont {Tsuneyuki}},\
  and\ \bibinfo {author} {\bibfnamefont {T.}~\bibnamefont {Schimizu}},\
  }\bibfield  {title} {\bibinfo {title} {{First-principles calculations of
  carrier-doping effects in SrTiO3}},\ }\bibfield  {journal} {\bibinfo
  {journal} {Physical Review B - Condensed Matter and Materials Physics}\
  }\href {https://doi.org/10.1103/PhysRevB.68.174107}
  {10.1103/PhysRevB.68.174107} (\bibinfo {year} {2003})\BibitemShut {NoStop}%
\bibitem [{\citenamefont {Goldschmidt}(1926)}]{Goldschmidt1926}%
  \BibitemOpen
  \bibfield  {author} {\bibinfo {author} {\bibfnamefont {V.~M.}\ \bibnamefont
  {Goldschmidt}},\ }\bibfield  {title} {\bibinfo {title} {{Die Gesetze der
  Krystallochemie}},\ }\bibfield  {journal} {\bibinfo  {journal} {Die
  Naturwissenschaften}\ }\href {https://doi.org/10.1007/BF01507527}
  {10.1007/BF01507527} (\bibinfo {year} {1926})\BibitemShut {NoStop}%
\bibitem [{\citenamefont {Mohanta}\ \emph {et~al.}(2021)\citenamefont
  {Mohanta}, \citenamefont {Ok}, \citenamefont {Zhang}, \citenamefont {Miao},
  \citenamefont {Dagotto}, \citenamefont {Lee},\ and\ \citenamefont
  {Okamoto}}]{Mohanta2021}%
  \BibitemOpen
  \bibfield  {author} {\bibinfo {author} {\bibfnamefont {N.}~\bibnamefont
  {Mohanta}}, \bibinfo {author} {\bibfnamefont {J.~M.}\ \bibnamefont {Ok}},
  \bibinfo {author} {\bibfnamefont {J.}~\bibnamefont {Zhang}}, \bibinfo
  {author} {\bibfnamefont {H.}~\bibnamefont {Miao}}, \bibinfo {author}
  {\bibfnamefont {E.}~\bibnamefont {Dagotto}}, \bibinfo {author} {\bibfnamefont
  {H.~N.}\ \bibnamefont {Lee}},\ and\ \bibinfo {author} {\bibfnamefont
  {S.}~\bibnamefont {Okamoto}},\ }\bibfield  {title} {\bibinfo {title}
  {{Semi-Dirac and Weyl fermions in transition metal oxides}},\ }\bibfield
  {journal} {\bibinfo  {journal} {Physical Review B}\ }\textbf {\bibinfo
  {volume} {104}},\ \href {https://doi.org/10.1103/PhysRevB.104.235121}
  {10.1103/PhysRevB.104.235121} (\bibinfo {year} {2021})\BibitemShut {NoStop}%
\bibitem [{\citenamefont {Rowe}(2018)}]{Rowe2018}%
  \BibitemOpen
  \bibfield  {author} {\bibinfo {author} {\bibfnamefont {D.~M.}\ \bibnamefont
  {Rowe}},\ }\href {https://doi.org/10.1201/9781420049718} {\emph {\bibinfo
  {title} {CRC Handbook of Thermoelectrics}}}\ (\bibinfo  {publisher} {CRC
  Press},\ \bibinfo {year} {2018})\BibitemShut {NoStop}%
\bibitem [{\citenamefont {Wei}\ \emph {et~al.}(2009)\citenamefont {Wei},
  \citenamefont {Bao}, \citenamefont {Pu}, \citenamefont {Lau},\ and\
  \citenamefont {Shi}}]{Wei2009}%
  \BibitemOpen
  \bibfield  {author} {\bibinfo {author} {\bibfnamefont {P.}~\bibnamefont
  {Wei}}, \bibinfo {author} {\bibfnamefont {W.}~\bibnamefont {Bao}}, \bibinfo
  {author} {\bibfnamefont {Y.}~\bibnamefont {Pu}}, \bibinfo {author}
  {\bibfnamefont {C.~N.}\ \bibnamefont {Lau}},\ and\ \bibinfo {author}
  {\bibfnamefont {J.}~\bibnamefont {Shi}},\ }\bibfield  {title} {\bibinfo
  {title} {{Anomalous thermoelectric transport of dirac particles in
  graphene}},\ }\bibfield  {journal} {\bibinfo  {journal} {Physical Review
  Letters}\ }\textbf {\bibinfo {volume} {102}},\ \href
  {https://doi.org/10.1103/PhysRevLett.102.166808}
  {10.1103/PhysRevLett.102.166808} (\bibinfo {year} {2009})\BibitemShut
  {NoStop}%
\bibitem [{\citenamefont {Zhu}\ \emph {et~al.}(2018)\citenamefont {Zhu},
  \citenamefont {Trevisanutto}, \citenamefont {Asmara}, \citenamefont {Xu},
  \citenamefont {Feng},\ and\ \citenamefont {Rusydi}}]{Zhu2018}%
  \BibitemOpen
  \bibfield  {author} {\bibinfo {author} {\bibfnamefont {T.}~\bibnamefont
  {Zhu}}, \bibinfo {author} {\bibfnamefont {P.~E.}\ \bibnamefont
  {Trevisanutto}}, \bibinfo {author} {\bibfnamefont {T.~C.}\ \bibnamefont
  {Asmara}}, \bibinfo {author} {\bibfnamefont {L.}~\bibnamefont {Xu}}, \bibinfo
  {author} {\bibfnamefont {Y.~P.}\ \bibnamefont {Feng}},\ and\ \bibinfo
  {author} {\bibfnamefont {A.}~\bibnamefont {Rusydi}},\ }\bibfield  {title}
  {\bibinfo {title} {{Generation of multiple plasmons in strontium niobates
  mediated by local field effects}},\ }\bibfield  {journal} {\bibinfo
  {journal} {Physical Review B}\ }\textbf {\bibinfo {volume} {98}},\ \href
  {https://doi.org/10.1103/PhysRevB.98.235115} {10.1103/PhysRevB.98.235115}
  (\bibinfo {year} {2018})\BibitemShut {NoStop}%
\bibitem [{\citenamefont {Harl}\ \emph {et~al.}(2007)\citenamefont {Harl},
  \citenamefont {Kresse}, \citenamefont {Sun}, \citenamefont {Hohage},\ and\
  \citenamefont {Zeppenfeld}}]{Harl2007}%
  \BibitemOpen
  \bibfield  {author} {\bibinfo {author} {\bibfnamefont {J.}~\bibnamefont
  {Harl}}, \bibinfo {author} {\bibfnamefont {G.}~\bibnamefont {Kresse}},
  \bibinfo {author} {\bibfnamefont {L.~D.}\ \bibnamefont {Sun}}, \bibinfo
  {author} {\bibfnamefont {M.}~\bibnamefont {Hohage}},\ and\ \bibinfo {author}
  {\bibfnamefont {P.}~\bibnamefont {Zeppenfeld}},\ }\bibfield  {title}
  {\bibinfo {title} {{Ab initio reflectance difference spectra of the bare and
  adsorbate covered Cu(110) surfaces}},\ }\bibfield  {journal} {\bibinfo
  {journal} {Physical Review B - Condensed Matter and Materials Physics}\
  }\textbf {\bibinfo {volume} {76}},\ \href
  {https://doi.org/10.1103/PhysRevB.76.035436} {10.1103/PhysRevB.76.035436}
  (\bibinfo {year} {2007})\BibitemShut {NoStop}%
\end{thebibliography}
\end{document}